\documentclass[11pt,a4paper,epsfig]{article}

\usepackage{amsfonts}
\usepackage{graphicx}
\usepackage{cite}
\usepackage[greek,english]{babel}
\usepackage[colorlinks=true,linkcolor=blue]{hyperref}
\title{\bf Topological and non-topological kink families in non-linear $(\mathbb{S}^1\times \mathbb{S}^1)$-Sigma models}

\author{A. Alonso-Izquierdo$^{(a,b)}$, A.J. Balseyro Sebastian$^{(b)}$, M. A. Gonzalez Leon$^{(a,b)}$
\\ {\normalsize {\it $^{(a)}$ Departamento de Matematica
Aplicada}, {\it Universidad de Salamanca, SPAIN}}\\ {\normalsize {\it $^{(b)}$ IUFFyM}, {\it Universidad de Salamanca, SPAIN}}}

\date{}

\begin{document}

\maketitle

\begin{abstract}
In this paper we construct a family of Hamilton-Jacobi separable non-linear $\mathbb{S}^1\times\mathbb{S}^1$ Sigma models for which the kink variety can be analytically identified and for which the linear stability of the emerging kinks is ensured. Furthermore, a model with only one vacuum point is found, where all kinks are forced to be non-topological. The non-simply connectedness of the torus guarantees the global stability of all the non-topological kinks in these models. 
\end{abstract}

\section{Introduction}
 Topological defects are solutions of non-linear scalar field theories whose energy densities describe localized particles which cannot decay into vacuum because of topological constrains in the configuration space. This type of solutions are present in areas as diverse as optics \cite{Optics,Mollenauer2006,Schneider2004,Agrawall1995}, biochemistry \cite{biochemistry,Yakushevich2004}, cosmology \cite{cosmology,Kolb1990,Kibble1976,Vilenkin1994,Vachaspati2006} and physics of materials \cite{graphene,Bishop1980,Eschenfelder1981,Jona1993,Strukov}. This ubiquity makes them an incredibly powerful tool to describe several phenomena in nature, where non-linearity is present.    

When the space-time has only one spatial dimension topological defects receive the name of kinks. This work will be focusing on these solutions, which necessarily join minima of the potential of the field theory. Multiples strategies have been applied to try to identify them analytically. Searching for Hamilton-Jacobi separability for the analogue mechanical system has proven to be one of the most successful methods to find them \cite{Bazeia1995, Shifman1998, Ito1985, Alonso1998, Alonso2002, Afonso2007, Afonso2008, Afonso2009, Alonso2013}. Other approaches are also highly productive, as for instance the deformation methods developed by Bazeia \cite{Bazeia2006a, Bazeia2006b, Almeida2004,Bazeia2006c,Souza2009, Cruz2009, Chumbes2010, Bazeia2011}, where a kink is used as seed for obtaining kinks in new theories. These papers highlight the rich variety of structures and phenomena that can arise when the deformation techniques are applied and several applications in physics are discussed.

A particularly rich context where kinks are sought corresponds to the non-linear Sigma model. In this class of models, instead of considering independent fields in $\mathbb{R}^n$, these must satisfy a condition that forces them to be coordinates on a Riemannian manifold. Several applications have been found for these models. For instance, spin chain dynamics can be modeled by kinks in a $\mathbb{S}^2-$Sigma model \cite{Haldane1983,Alonso2008b,Alonso2008c,Alonso2010,Alonso1}. When a simply connected manifold like $\mathbb{S}^2$ is considered as target manifold, non-topological kinks can usually decay into vacuum. This is forbidden for certain types of loops when a non-simply connected manifold is chosen as target manifold, like the two-dimensional torus. The torus has been used before to enrich models in several contexts. In \cite{Vortex} an effective theory for vortex strings is constructed on the torus $\mathbb{S}^1 \times \mathbb{S}^1$ in the context of Skyrme models. In \cite{Garcia2023} solutions of the dynamics of a supermembrane compactified on the target space $M_9\times \mathbb{T}^2$ are obtained. In particular, by the use of an ansatz Q-ball-like solutions on the M2-brane with non-trivial worldvolume fluxes are found. These are also referred to as Q-torus in \cite{Garcia2023}. In \cite{Nematic} the equilibrium configurations of a thin surface of nematic liquid covering a torus are studied. In \cite{NonCommutative} a non-commutative version of the classical Sigma model is constructed by replacing the target space by a particular case of a $C^*-$algebra, a non-comutative torus. A non-commutative action functional is defined and it is shown that the corresponding Euler-Lagrange equations give rise to non-commutative versions of the classical field equations. In the present work we are interested in finding analytic solutions for Sigma models on the torus. In \cite{A} a family of non-linear $(\mathbb{S}^1\times \mathbb{S}^1)$-Sigma models on the torus that admit kink-type solutions is constructed. Among its solutions, non-contractible non-topological kinks are found for the case of models with two minima. Interestingly, the global stability of these kinks is ensured by the non-simply connectedness of the target manifold. However, that family of models does not include the scenario with only one minimum. This situation is extremely interesting since every kink in the model is forced to be non-topological. This work builds upon the research conducted in \cite{A} and further develops the established research line. In particular, another family of potentials for a non-linear $(\mathbb{S}^1\times \mathbb{S}^1)$-Sigma model is constructed, where apart from a rich kink variety for each member of this family, a model with only one minimum is found.

This research is structured as follows. In section $2$ the type of Sigma models on the torus $(\mathbb{S}^1\times \mathbb{S}^1)$ that shall be studied will be presented. In section $3$, a particular family of these models is chosen and its kink variety for certain members of this family is analyzed in section $4$. The number of minima will be decreasing until the case of only one minimum is obtained. Finally, conclusions are presented in section $5$.

\section{Non-linear $(\mathbb{S}^1 \times \mathbb{S}^1)$-Sigma models in (1+1)-dimensions}

We shall deal with non-linear Sigma models with the $2-$dimensional torus $\mathbb{S}^1 \times\mathbb{S}^1$ embedded in $\mathbb{R}^3$ as target space. This type of theory can be defined by two scalar fields that represent poloidal-toroidal coordinates on the torus. Therefore, they are the two scalar fields $\theta,\varphi: \mathbb{R}^{1,1} \rightarrow \mathbb{S}^1$ on the $(1+1)-$dimensional Minkowski space $\mathbb{R}^{1,1}$ related to Cartesian coordinates $\phi^1,\phi^2,\phi^3$ in $\mathbb{R}^3$ as
\begin{eqnarray}
\phi^1(\theta,\varphi) &=& (R+r\sin \theta)\cos \varphi \, ,  \nonumber \\
\phi^2(\theta,\varphi) &=& r \cos \theta \label{tor} \, ,\\
\phi^3(\theta,\varphi) &=& (R+r\sin \theta)\sin \varphi \nonumber 
\end{eqnarray}
with poloidal $\theta\in [0,2\pi)$ and toroidal $\varphi\in[0,2\pi)$ coordinates and where $R$ and $r$ are the major and minor radii of the torus $R>r>0$. The metric tensor in these coordinates has the form
\begin{equation}\label{eq:metrictensor}
    g= {\rm diag}\,\{r^2,(R+r\sin\theta)^2 \} \, ,
\end{equation}
 Let us also choose $\eta_{00}=-\eta_{11}=1$ and $\eta_{12}=\eta_{21}=0$ for the metric tensor on the Minkowski space $\mathbb{R}^{1,1}$. With these fields a non-linear Sigma model is constructed, for which the action reads
\begin{equation}
S[\theta, \varphi]= \int_{\mathbb{R}^{1,1}}  \, \left[ \frac{1}{2} r^2 \eta^{\mu \nu}\frac{\partial\theta}{\partial x^\nu}\frac{\partial\theta}{\partial x^\mu}+\frac{1}{2}(R+r\sin\theta)^2 \eta^{\mu \nu}\frac{\partial\varphi}{\partial x^\nu}\frac{\partial\varphi}{\partial x^\mu} -V(\theta,\varphi) \right] \, dx ~ dt \, , \label{action}
\end{equation}
where $V:\mathbb{S}^1 \times\mathbb{S}^1 \rightarrow \mathbb{R}$ is a non-negative \textit{potential function} and Einstein summation convention is employed for the space-time indices. This action, in turn, will lead us to the following field equations
\begin{equation}
\eta^{\mu\nu} \frac{\partial^2 u^i}{\partial x^\mu  \partial x^\nu} + \eta^{\mu\nu}\Gamma_{jk}^i \frac{\partial u^j}{\partial x^\mu} \frac{\partial u^k}{\partial x^\nu}+ g^{ij} \frac{\partial V}{\partial u^j}=0  \, , \quad u^1=\theta \, ,\, u^2=\varphi \, \, \, , \, \, \, i,j=1,2 \, . \label{fenlv2}
\end{equation}
In the generic expression (\ref{fenlv2}) Einstein summation convention is used for both latin and greek indices and $\Gamma_{jk}^i$ denotes the Christoffel symbols associated to the metric tensor $g$ defined in (\ref{eq:metrictensor})
\begin{equation}
    \Gamma^1_{11}=\Gamma^1_{12}=\Gamma^1_{21} = \Gamma^2_{11}= \Gamma^2_{22}=0, \hspace{0.5cm} \Gamma^1_{22}=-\frac{1}{r} \cos\theta (R+r\sin \theta),\hspace{0.5cm} \Gamma^2_{12} =\Gamma^2_{21}= \frac{r\cos \theta}{R+r\sin \theta} \, .
\end{equation}
Solutions $\Sigma(t,x) \equiv (\theta(t,x),\varphi(t,x))$ of the field equations (\ref{fenlv2}) must belong to the configuration space defined by finite energy
\begin{eqnarray*}
E[\theta, \varphi]&=& \int^{\infty}_{-\infty}  \, \left\{\frac{1}{2} r^2 \left[\left(\frac{\partial\theta}{\partial t}\right)^2+\left(\frac{\partial\theta}{\partial x}\right)^2\right]+\frac{1}{2}(R+r\sin\theta)^2 \left[\left(\frac{\partial\varphi}{\partial t}\right)^2+\left(\frac{\partial\varphi}{\partial x}\right)^2\right] +V\left[\theta,\varphi\right]\right\} \, dx \, .
\end{eqnarray*}
 This condition makes these solutions tend asymptotically to a zero of the potential function $V(\theta,\varphi)$ and their derivatives to zero at the ends of the spatial line. Let us denote ${\cal M}$ the set of zeroes of the potential $V(\theta,\varphi)$
\[
{\cal M} = \{v_j =(\theta_j,\varphi_j) \in \mathbb{S}^1 \times \mathbb{S}^1 \, |\, \,  V(\theta_j,\varphi_j) =0, \,\, j=1,2,\dots \}\, .
\]
which shall be referred to as vacua of the theory. Solutions of field equations that travel with constant velocity will be of particular interest, given their interpretation as moving extended particles. However, since these field theories are Lorentz invariant, static solutions are sought instead. For these solutions the energy is of the form
\[
    E[\theta(x), \varphi(x)]=\displaystyle\int_{-\infty}^{\infty}\varepsilon(x) dx
\]
where $\varepsilon(x)$ is the energy density for a given static solution $(\theta(x), \varphi(x))$.

\section{A family of models with different number of vacua}

The next step to construct a Sigma model, as described in the previous section, is to define the potential function on $\mathbb{S}^1\times \mathbb{S}^1$. In this paper the set of potential functions
\begin{equation}\label{eq:Potgeneral}
V_{n_1,n_2}(\theta,\varphi)= \frac{1}{2} \Big[ \frac{m_1^2 n_1^2}{r^2} \cos^2(n_1\theta) +\frac{m_2^2 n_2^2}{(R+r \sin (\theta))^2} \cos^2 (n_2\varphi) \Big] \, ,  \label{potential}
\end{equation}
 is proposed, where $m_1,m_2$ are positive real numbers and $n_1$ and  $n_2$ are integers or half-integers so that these potential functions are periodic on the torus. Furthermore, the set of vacua ${\cal M}$ of each potential contains a total of $|4 n_1 n_2|$ points distributed on the torus
\[
{\cal M}_{n_1,n_2}= \Big\{ \Sigma_{k_1,k_2} = (\theta_{k_2}, \varphi_{k_1})= \Big(\frac{\pi}{2n_1}+\frac{k_1 \, \pi}{n_1} \, , \, \frac{\pi}{2n_2} +\frac{k_2 \, \pi}{n_2}  \Big) \quad | \quad k_i \in \mathbb{Z} \, \Big\}\, , 
\]
\begin{figure}[ht]
\begin{center}
    \includegraphics[height=4.5cm]{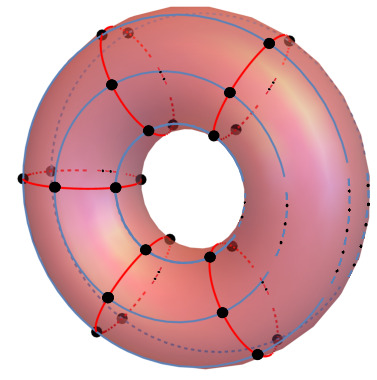} \hspace{0.5cm}
    \includegraphics[height=4.5cm]{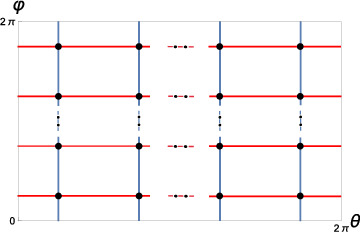}
    \caption{\small Set of vacua for an arbitrary member of this family of potentials (\ref{potential}). While in the $\theta-\varphi-$plane an infinite number of minima appear, on the torus these correspond to a finite number of vacua for fixed $n_1$ and $n_2$.}\label{fig:16Vacios}
    \end{center}
\end{figure}
which is exactly the same number of regions on the torus. Cases of (\ref{eq:Potgeneral}) with four, two and one vacua will be thoroughly explored, where globally stable kinks are expected to be identified. For potentials that are derived from a superpotential $W:\mathbb{S}^1 \times\mathbb{S}^1\rightarrow\mathbb{R}$ as follows 
\begin{equation}
V(u^1,u^2)= \frac{1}{2} g^{ij}\, \frac{\partial W}{\partial u^i} \frac{\partial W}{\partial u^j} \, ,\hspace{1cm} u^1=\theta \, ,\, u^2=\varphi \, \, \, , \, \, \, i,j=1,2 \, ,  \label{ps}
\end{equation}
making use of the Bogomol'nyi arrangement the static energy is minimized when the following system of differential equations hold
\begin{equation}
\frac{du^i}{dx}= g^{ij} \, \frac{\partial W}{\partial u^j} \, ,\hspace{1cm} u^1=\theta \, ,\, u^2=\varphi \, \, \, , \, \, \, i,j=1,2. \label{bpsb}  
\end{equation}
Notice how superpotentials come always at least in pairs, since $W\rightarrow -W$ produces the same potential function. Furthermore, for this type of potentials the energy of the static solutions depends exclusively on which minima the kink is connecting 
\begin{equation}\label{eq:EnergyBPS}
   E=\Big\vert \lim_{x\rightarrow \infty} W[\theta(x),\varphi(x)]- \lim_{x\rightarrow -\infty}W[\theta(x),\varphi(x)]\Big\vert \,  
\end{equation}
when $W$ is a differentiable function. For the family of potentials (\ref{eq:Potgeneral}) four families of differentiable superpotentials can be found
\begin{equation}
W_{n_1,n_2}(\theta,\varphi)= (-1)^{\epsilon_1} m_1  \sin (n_1\theta) + (-1)^{\epsilon_2} m_2  \sin (n_2\varphi) \, ,  \label{superpotential}
\end{equation}
with $\epsilon_1, \epsilon_2=0,1$. This allows to write Bogomol'nyi equations (\ref{bpsb}) as 
\begin{equation}
\frac{d \theta}{dx} = (-1)^{\epsilon_1} \, \frac{m_1 \, n_1}{r^2} \cos (n_1\theta) \, ,\hspace{0.5cm} \frac{d \varphi}{dx} = (-1)^{\epsilon_2} \,  \frac{m_2 \, n_2 }{(R+r\sin \theta)^2} \cos (n_2\varphi) \, . \label{eq:PrimerOrden}
\end{equation}
Notice that since $W$ is a bounded function the energy (\ref{eq:EnergyBPS}) of any solution of these equations is finite. It is worth noting that the potential functions (\ref{potential}) exhibit the symmetries $\varphi\rightarrow\varphi +\frac{k_2}{n_2}\pi$ for any $k_2\in \mathbb{Z}$. As expected, these are also present in Bogomol'nyi equations. However, transformations $\theta\rightarrow\theta +\frac{k_1}{n_1}\pi$ with $k_1\in \mathbb{Z}$ also allows us to identify solutions in different pairs of Bogomol'nyi equations. If $(\theta_s,\varphi_s)$ is a solution for a given choice of $\epsilon_1$ and $\epsilon_1$, then the following identification can be made
\begin{equation}
    \left(\theta_s,\varphi_s,\epsilon_1,\epsilon_2\right) \longleftrightarrow \left(\theta_s +\frac{k_2}{n_2}\pi ,\varphi_s,(-1)^{k_1}\epsilon_1,\epsilon_2\right)
\end{equation}
That is, solutions that belong to the different four families of superpotentials will also be related. These two types of identifications combined, both in $\theta$ and in $\varphi$, will split the torus into $|4n_1 n_2|$ regions where the kink variety will be replicated.

As it is well known, topological kinks can exhibit global stability since they asymptotically connect different vacua. This means that their topological sector is different from that of vacua. On the other hand, non-topological kinks can also be globally stable if the target space presents certain topological properties. This type of solutions shall be referred to as ``vrochoson'', which comes from the ancient Greek word ``\textgreek{βροχος}'' (vrochos), meaning ``loop'' or ``ring''. These solutions are different from the recently baptized ``lavions'', which are locally (but not globally) stable non-topological kinks, see \cite{Halcrow,Alonso5}, due to the existence of singularities in the potential. One example of vrochoson can be found in the non-linear $(\mathbb{S}^1\times \mathbb{S}^1)$-Sigma model investigated in \cite{A,Alonso4}, where even if kinks asymptotically connect a minimum with itself the topology of the target space prevents them from decaying into vacuum. In this work a different family of potentials has been constructed so that also a model with only one vacuum point is found. Once again, given the non-simply connnectedness of the target manifold, non-topological kinks on the torus will be globally stable.
Moreover, unlike the previous family of potentials, the new one will be generated from a family of separable superpotentials, which will produce not one but two families of kinks in each model. Notice that since the potential is separable in these coordinates, the Bogomol'nyi arrangement for the potential family (\ref{potential}) can be made by these four superpotentials. Bogomol'nyi equations can be solved analytically and their solutions will be described in the following section.

\section{Singular kinks and families of energy degenerate kinks}
Solutions of Bogomol'nyi equations will be classified depending on the number of coordinates that remain constant. This leaves us with four types of solutions. The first one is comprised by constant solutions in both coordinates $\theta$ and $\varphi$, that is, values that correspond to the set of vacua of the system, which have been mentioned in Section $3$. The second and third types correspond to orbits with $\theta$ or $\varphi$ constant respectively, which shall be referred to as singular kinks. Lastly, the fourth one contains families of solutions with neither $\theta$ nor $\varphi$ constant: 
\begin{enumerate}
        \item Singular $\Phi$-kinks: The $|2n_1|$ \emph{toroidal} trajectories defined by $\theta=\frac{\pi}{2 n_1}+\frac{k_1}{n_1} \pi$ that cross at least one minimum will be referred as $\Phi$-kink orbits. First order equations (\ref{eq:PrimerOrden}) with this condition explicitly provide these solutions $\Phi_K^{(k_1,k_2)}(x) = (\theta_K(x),\varphi_K(x))$ 
    \begin{equation}
      \hspace{-0.38cm} \Phi_K^{(k_1,k_2)}(x) = \left( \frac{2k_1+1}{2 n_1} \pi \,\, , \,\, \frac{k_2+1}{n_2} \pi + \frac{1}{n_2}{\rm Gd} \left[(-1)^{k_2+\epsilon_2+1} \, n_2^2 \, A_{k_1} \bar{x})\right]\right)\, ,
        \label{eq:SingularPhi}
    \end{equation}
    with $\bar{x}=x-x_0$ where $x_0$ can be considered the kink center, ${\rm Gd}$ denotes the Gudermannian function ${\rm Gd}[z]=-\frac{\pi}{2}+2 \arctan{e^z}$ and
    $$A_{k_1}=\frac{m_2}{(R+r \sin{\frac{2k_1+1}{2n_1}\pi})^2}\, .$$
    In this case, kinks revolve around the torus center, connecting minima located at $\varphi=\frac{2k_1+1}{2 n_1} \pi$ and $\varphi=\frac{2k_1+3}{2 n_1}\pi$. Lastly, different energy density profiles associated with solutions (\ref{eq:SingularPhi}) will appear depending on $k_1$
    \begin{equation}
        \varepsilon\left(\Phi_K^{(k_1,k_2)}(x)\right)=\frac{m_2^2 n_2^2}{\left(R + r \sin{\left(\frac{2k_1+1}{2 n_1}\pi\right)}\right)^2} ~ {\rm sech}^2\frac{n_2^2 m_2 \bar{x}}{\left(R + r \sin{\left(\frac{2k_1+1}{2 n_1}\pi\right)}\right)^2} \, ,
    \end{equation}
    which describe different single lumps with the same total energy
    \begin{equation}
    E[\Phi_K^{(k_1,k_2)}] =E[\Phi_K^{[k_1]}] =2 m_2 \, . \label{ener1}
    \end{equation}
    Their kink profiles depend on $k_1$ and $k_2$, but the energy density profile of all these singular kinks depend only on $k_1$. Taking this into consideration notation $\Phi_K^{[k_1]}$ with $[k_1]\equiv[k_1]_{|2n_1|}$ is employed, where the identification $[k_1]=[k_1+2\pi]$ is made. Notice that only $n_1$ or $n_1+1$ different energy density profiles emerge for even and odd $n_1$ respectively.
    
     \item Singular $\Theta$-kinks: The $|2n_2|$ \emph{poloidal} kink orbits that cross any vacuum $v_i$ defined by the condition $\varphi=\frac{\pi}{2 n_2}+\frac{k_2}{n_2} \pi$ (orthogonal to toroidal trajectories) will be referred as singular $\Theta$-kink orbits. Equations (\ref{eq:PrimerOrden}) with this condition produce their kink profile $\Theta_K^{(k_1,k_2)}(x)= (\theta_K(x),\varphi_K(x))$
     \begin{equation}
     \hspace{-0.44cm}
	\Theta_K^{(k_1,k_2)}(x) = \left( \frac{k_1+1}{n_1}\pi +\frac{1}{n_1}{\rm Gd}\left[ \frac{(-1)^{k_1+\epsilon_1+1}\, m_1 n_1^2}{r^2} \,\bar{x}\right],\frac{2k_2+1}{2n_2}\pi\right)\, ,
\label{eq:SingularTheta}
     \end{equation}
    whose total energy is independent of any $k_i$ and only one energy density profile, which corresponds again to a single lump, will be obtained
    \[
    E[\Theta_K^{(k_1,k_2)}]= E[\Theta_K]=2 m_1 \, , \qquad \varepsilon\left(\Theta_K^{(k_1,k_2)}\right)=\frac{m_1^2 n_1^2}{r^2} ~ {\rm sech}^2\left(\frac{n_1^2 m_1\bar{x}}{r^2}\right) \, ,
    \]
    for all $k_1,k_2$. Once more, given that the energy density is independent of $k_1$ and $k_2$ notation $\Theta_K$ is used. These solutions describe only one extended particle. Notice that for both $\Phi_K$ and $\Theta_K-$kinks, parameters $m_2$ and $m_1$ respectively modulate not only their energy but also the maximum height of the peak in their energy density.
    
    \item Families of energy degenerate kinks: In this case, equations (\ref{eq:PrimerOrden}) can also be analytically solved when neither of the variables are constant 
    \begin{eqnarray*}
\theta(x) &=& \frac{k_1+1}{n_1}\pi +\frac{1}{n_1}{\rm Gd}\left[(-1)^{\epsilon_1+k_1+1} \frac{m_1 n_1^2}{r^2}(x-x_0)\right] \, ,\\
\varphi(x) &=& \frac{k_2+1}{n_2}\pi +\frac{1}{n_2}{\rm Gd}\left[(-1)^{\epsilon_2+k_2+1} m_2 n_2^2(\xi(x)-\xi_0)\right] \, , 
\end{eqnarray*}
where $\xi_0\in\mathbb{R}$ and $\xi(x)$ is a new spatial parameter defined by
\begin{equation}\label{eq:parameter}
    \xi(x)=\displaystyle\int\frac{dx}{(R+r\sin{\theta(x)})^2}\, .
\end{equation}
 Notice that $x_0$ represents a shift in $x$, while $\xi_0$ is a parameter that changes the orbit of the solution. On the other hand, from (\ref{eq:PrimerOrden}) the kink orbit flow in the phase plane can be determined by the equation
\begin{equation}\label{eq:OrbitEquation}
   \frac{d\theta}{d\varphi} = (-1)^{\epsilon} \, \frac{m_1 \, n_1}{m_2 \, n_2} \frac{(R+r\sin (\theta))^2}{r^2} \frac{\cos (n_1\theta)}{\cos (n_2\varphi)} \, , \label{eq:orbita}
\end{equation}
where $\epsilon=\epsilon_1-\epsilon_2$ and the integration constant that will appear when integrating the orbit equation will have the same role as $\xi_0$ in discriminating members in these families. It is clear from the expression above that two families of orbits, denoted as $\Sigma_K^{(k_1,k_2,\epsilon)} (x;\gamma)$ are found, one for each value of $\epsilon$ modulo $2$ and where $\gamma$ will be the integration constant. Notice that the family label can change $\epsilon\rightarrow \epsilon+1$ if one of the $\epsilon_i\rightarrow \epsilon_i+1$ is shifted. This transformation can be performed by applying to a superpotential (\ref{superpotential}) one of the following transformations:
\begin{equation}
    (\theta,\varphi)\rightarrow \left(\theta,\varphi+\frac{2l+1}{n_2}\pi\right) \, , \qquad (\theta,\varphi)\rightarrow \left(\theta,\frac{2\pi l}{n_2}-\varphi\right) \, , 
\end{equation}
for $l\in \mathbb{Z}$, so that another of the four available superpotentials is obtained and it is a symmetry of the potential function. These symmetries can relate different families of solutions, which may even belong to different topological sectors. On the other hand, note that if $n_i=\frac{1}{2}$ the shift in the corresponding variables that relates both families is $2\pi$. This can be seen in the poloidal-toroidal plane when more copies of the same torus are represented. Moreover, the kink energy (\ref{eq:EnergyBPS}) will be the same for any member of a family since the initial and final points coincide, but also, employing the symmetries described above, it can be easily proved that members of even different families share the same energy and it is independent of $k_1$ and $k_2$. In fact, it is solely modulated by parameters $m_1$ and $m_2$, giving place to the energy sum rule: 
\begin{equation}\label{eq:EnergySumRule}
    E\left[\Sigma_K^{(k_1,k_2,0)}\right]=E\left[\Sigma_K^{(k_1,k_2,1)}\right]=2(m_1+m_2)=E[\Phi_K^{(k_1,k_2)}]+E[\Theta_K^{(k_1,k_2)}] \, .
\end{equation} 
    \end{enumerate}
    
    In principle, parameters $n_1$ and $n_2$ are allowed to be negative, but the sign that appears in reflections $n_i\rightarrow -n_i$ in any or both cases $i=1,2$ can be absorbed by $\theta \rightarrow - \theta$ or/and $\varphi \rightarrow -\varphi$. Consequently, only positive $n_i$ will be considered from now on, since symmetries will generate the negative cases. The kink variety for the different members of the family of potentials (\ref{eq:Potgeneral}) with different number of minima will be very different. Notice also that, unlike in simply connected spaces, labeling the minima that the kink is asymptotically joining is not enough to distinguish between topological sectors. Indeed, even if these two minima are fixed, kinks revolving a different number of times around a direction on the torus will not belong to the same topological class. Now, since an infinite number of topological sector arise, these will be grouped into ``topological clusters'' in which solutions link the same minima. The term topological is then used for kinks that belong to clusters in which solutions connect different minima and non-topological for kinks that belong to clusters that connect a minimum with itself. The following four particular cases of the potential (\ref{potential}) are chosen to be studied in detail:
    
    \begin{itemize}
    
        \item Case $1$: Aiming for four minima the potential with values $n_1=n_2=1$ is chosen as representative. This implies the existence of sixteen disjoint topological clusters in the configuration space, but only twelve of them will contain kinks. Eight singular kinks and antikinks and eight families of kinks will be found. 
        
    	\item Case $2$: Values $n_1=1$ and $n_2=\frac{1}{2}$ in the potential (\ref{potential}) make the number of minima decrease down to two. Four topological clusters containing kinks appear. Two topological and two non-topological singular kinks and antikinks and two families of topological kinks are identified. These non-topological kinks will be non-contractible to points.

    	\item Case $3$: If the values of the parameters are swapped with respect to Case $2$, $n_1=\frac{1}{2}$ and $n_2=1$, another potential with two minima is obtained. The same number of singular kinks and antikinks are found and two families of topological kinks are found as well, although its structure is different.
    	
    	\item Case $4$: Finally, when $n_1=n_2=\frac{1}{2}$ every kink must be non-topological, since there is only one minimum. Only one topological cluster can exist and it will contain only non-contractible kinks. Indeed, two non-topological singular kinks and antikinks and one family of non-topological kinks on the torus are found. 
    \end{itemize}
Solutions of Bogomol'nyi equations for a given superpotential minimize the energy in the corresponding topological sectors. The fact that the Bogomol'nyi arrangement can only be performed by one smooth superpotential guarantees the linear stability of all these kinks. Indeed, there is no other superpotential for which the energy of kinks in a topological sector could be lower.

\subsection{Kink variety for a model with four vacua}

Case $1$ involves a potential function with four minima since $|4 n_1 n_2|=4$. Among the existing three positive possibilities, $n_1=\frac{1}{2}$ and $n_2=2$, $n_1=2$ and $n_2=\frac{1}{2}$ and $n_1=1$ and $n_2=1$, the latter will be chosen to be studied. A similar analysis can be performed for the rest of the cases. The potential function (\ref{potential}), which has the form in this case
\begin{equation}
V_{1,1}(\theta,\varphi)= \frac{1}{2} \left[\frac{m_1^2}{r^2} \, \cos^2\theta +\frac{m_2^2}{(R+r \sin \theta)^2} \, \cos^2 \varphi\right]  \, , \label{potentialZ}
\end{equation}
has four minima symmetrically distributed on the torus
\begin{eqnarray*}
 {\cal M}_{1,1} & = & \textstyle \{ v^1=(\frac{\pi}{2},\frac{\pi}{2}); v^2=(\frac{3\pi}{2},\frac{\pi}{2}) ; v^3=(\frac{3\pi}{2},\frac{3\pi}{2}) ; v^4=(\frac{\pi}{2},\frac{3\pi}{2})\} \, ,
\end{eqnarray*}
see Figure \ref{figure:A1y2}. The four superpotentials that this potential function admits are periodic on the torus and read
\begin{equation}
W_{1,1}(\theta,\varphi)=(-1)^{\epsilon_1} \, m_1 \, \sin \theta + (-1)^{\epsilon_2} \, m_2 \, \sin \varphi \, .  \label{superpotentialA}
\end{equation}
Translations $\theta \to \theta+\pi$ and $\varphi \to \varphi+\pi$ relate solutions from superpotentials with different $\epsilon_1$ and $\epsilon_2$, even if only translation $\varphi \to \varphi+\pi$ is a symmetry of the potential $V_{1,1}\left(\theta,\varphi\right)$. This implies that any orbit is replicated in other three regions of the torus. The torus is therefore split into four regions where the structure of the kink variety is identical. 
Bogomol'nyi equations in this case provide four singular topological kinks and antikinks as well as two families of topological kinks in each region:  
\begin{itemize}
	\item Singular $\Phi$-kinks: General expression (\ref{eq:SingularPhi}) provides solutions that link minimum $v^1$ to $v^4$ and minimum $v^2$ to $v^3$ 
\begin{equation}
\Phi_K^{(k_1,k_2)} (x)= \left(\frac{2k_1+1}{2} \pi\,\, , \,\, (k_2+1) \pi+ ~ {\rm Gd}\left[(-1)^{\epsilon_2+k_2+1}\frac{m_2 \bar{x}}{(R+(-1)^{k_1}r)^2}\right] \, \right) \, .
\end{equation}
Employing notation $\Phi_K^{[k_1]}$, the $\Phi_K^{[0]}-$kinks will correspond to exterior singular kinks on the torus while $\Phi_K^{[1]}-$kinks to interior kinks, see Figure \ref{figure:A1y2}. Label $k_2=0,1$ will distinguish beetween the two pieces of each circumference. As expected, the energy density is concentrated around a single point. However, even if the energy of all these singular kinks is the same 
$E[\Phi_K]= 2m_2$, two different energy density profiles arise depending on the label $[k_1]=0,1$
    \begin{equation}
        \varepsilon\left(\Phi_K(x)\right)=\frac{m_2^2}{\left(R +  (-1)^{k_1} r \right)^2} ~ {\rm sech}^2\left(\frac{m_2 \bar{x}}{\left(R +  (-1)^{k_1} r \right)^2}\right) \, ,
    \end{equation}
which means that there are two different types of extended particles. The peak in the energy density of kinks with $[k_1]=1$ will be more pronounced.  
\begin{figure}[ht]
\centerline{\includegraphics[height=4.5cm]{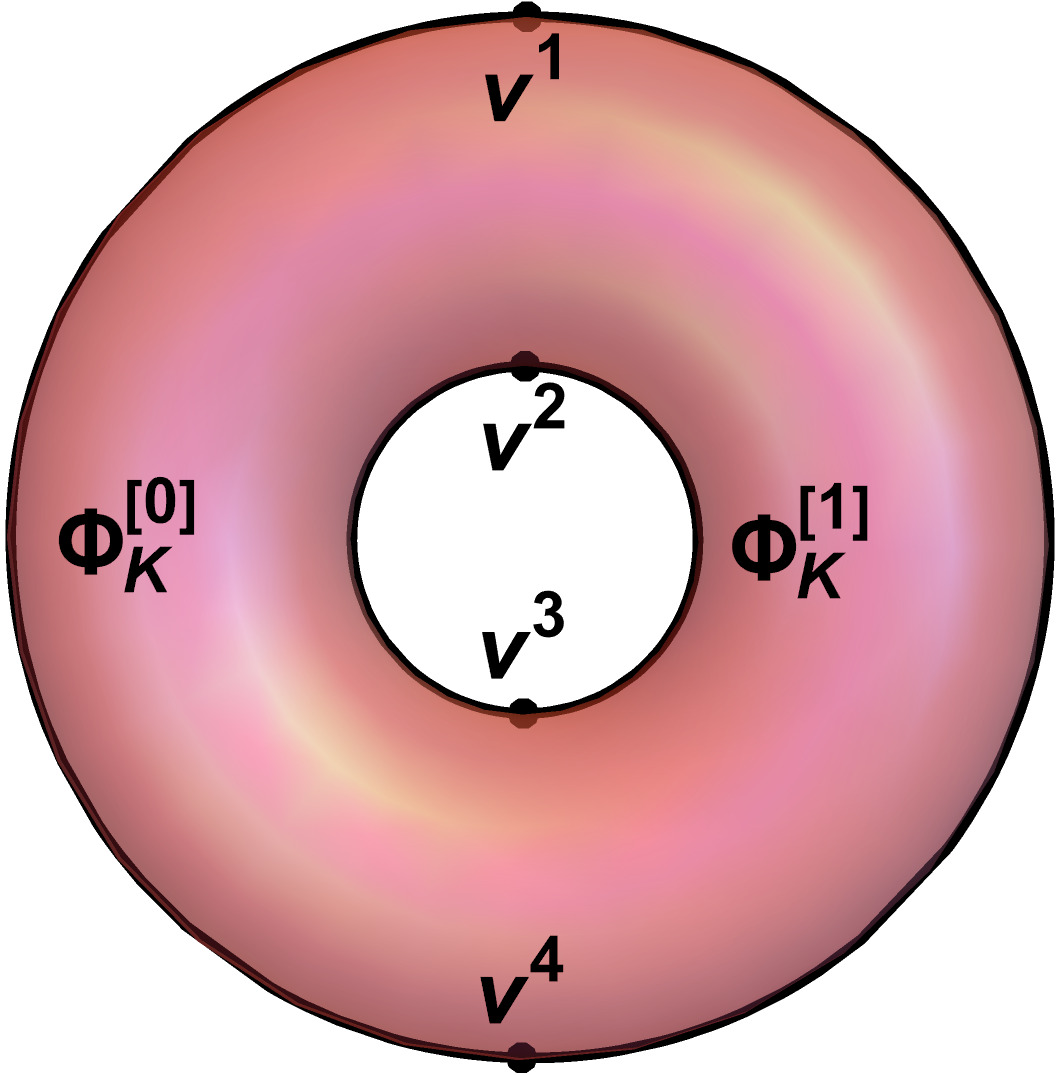} \hspace{0.5cm}
\includegraphics[height=4.5cm]{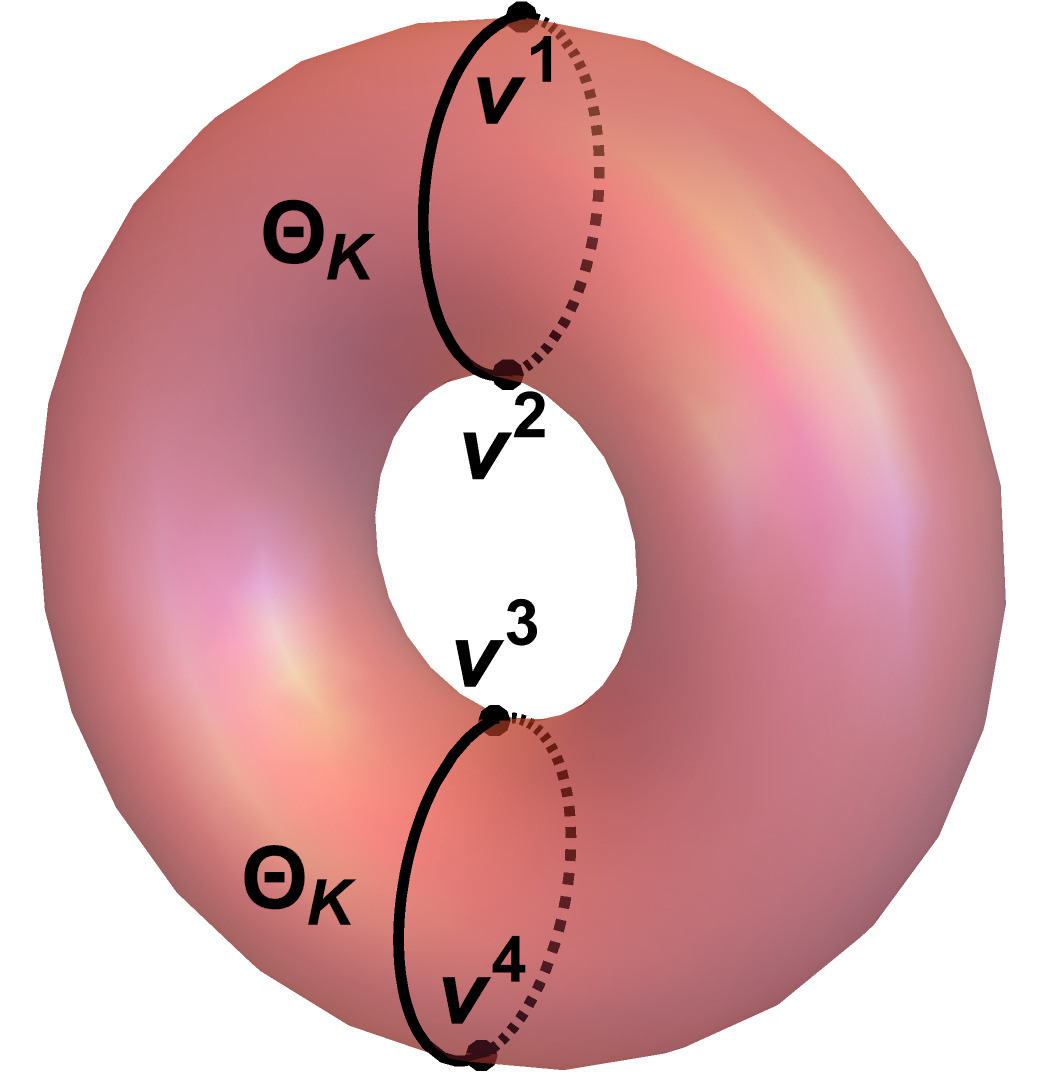}}
\caption{\small $\Phi_K$ and $\Theta_K-$kink orbits are represented on the torus, where dotted lines show kinks on the back side. See how these singular kinks combined split the torus into four regions related by symmetries of the potential.}
\label{figure:A1y2}
\end{figure}

\item Singular $\Theta$-kinks: Four topological kinks and antikinks in the poloidal direction are obtained when $\varphi$-constant orbits are sought in the first order equations (\ref{eq:PrimerOrden})
\begin{equation}
\Theta_K^{(k_1,k_2)} (x) =  \left((k_1+1) \pi+ {\rm Gd}\left[(-1)^{\epsilon_1+k_1+1}\frac{m_1 \bar{x}}{r^2}\right]\hspace{0.2cm} , \hspace{0.2cm} \frac{2 k_2+1}{2} \pi\right) \, ,
\end{equation}
where $k_1=0,1$ and $k_2=0,1$ are enough to represent all these kinks on the torus. While $\Theta^{(k_1,0)}$-kinks connect points $v^1$ and $v^2$, $\Theta^{(k_1,1)}$-kinks join minima $v^3$ and $v^4$, see orbits in Figure \ref{figure:A1y2}. This means that $k_2$ selects between the upper and lower circumference and $k_1$ between the front and back side of the torus. All $\Theta_K(x)$-kinks share the same energy density profile and energy:
\[
\varepsilon\left(\Theta_K(x)\right)=\frac{m_1^2}{r^2} ~ {\rm sech}^2\left(\frac{m_1 \bar{x}}{r^2}\right) \, , \qquad E[\Theta_K]=2m_1 \, .
\]
\item Kink families: The two possible values of $\epsilon$ in (\ref{eq:orbita}) give rise to two families of kinks in each region on the torus. Even if it is possible to obtain the explicit form of the parameter (\ref{eq:parameter}) for the families when $n_1=1$, due to its length it is omitted. Instead, let us give explicitly the kink orbits by solving equation (\ref{eq:OrbitEquation}), whose solution in this case reads
\begin{equation}
\sin \varphi=  \tanh \left[\gamma + (-1)^{\epsilon} \, \frac{m_2}{2 m_1} \frac{r^2}{(R+r)^2} \ln \left|\frac{(1+ \sin \theta)^{\left(\frac{R+r}{R-r}\right)^2} e^{\left(\frac{R+r}{R-r} \frac{2r}{R+r\sin \theta}\right)}}{(1-\sin \theta)(R+r \sin \theta)^{\frac{4 r R}{(R-r)^2}}}\right|\right] \, ,
\label{eq:orbA}
\end{equation}
where the parameter $\gamma\in\left(-\infty,\infty\right)$ is an integration constant that labels every kink member $\Sigma_K^{(k_1,k_2,\epsilon)} (x;\gamma)$ in each family. Let us briefly discuss both families, which, as stated before, have the same energy:
\begin{itemize}
    \item Kink family with $\epsilon=0$: Minima $v^1$ and $v^3$ are connected in the four equivalent regions of the torus by curves that densely fill each region, see Figure \ref{4vFam1}. The limit when $\gamma\rightarrow -\infty$ in the orbit equation leads either to $\varphi=\frac{3\pi}{2}+2 \pi l_1$ with $l_1\in\mathbb{Z}$ or to $\theta=\frac{\pi}{2}+2 \pi l_2$ with $l_2\in\mathbb{Z}$. These correspond to a $\Phi^{[0]}_K-$kink and a $\Theta_K-$kink, which will be different for different regions labeled by $k_1$ and $k_2$. On the other hand, the limit $\gamma\rightarrow \infty$ forces either $\varphi=\frac{\pi}{2}+2 \pi l_1$ or $\theta=\frac{3\pi}{2}+2 \pi l_2$. These correspond to different $\Theta_K-$kinks and $\Phi^{[1]}_K-$kinks.    
    \begin{figure}[h]
\centerline{\includegraphics[height=4.8cm]{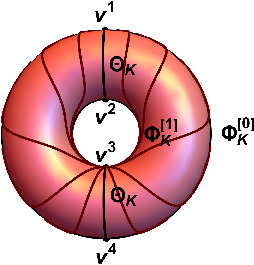}  \hspace{0.5cm} \includegraphics[width=6cm,height=4.5cm]{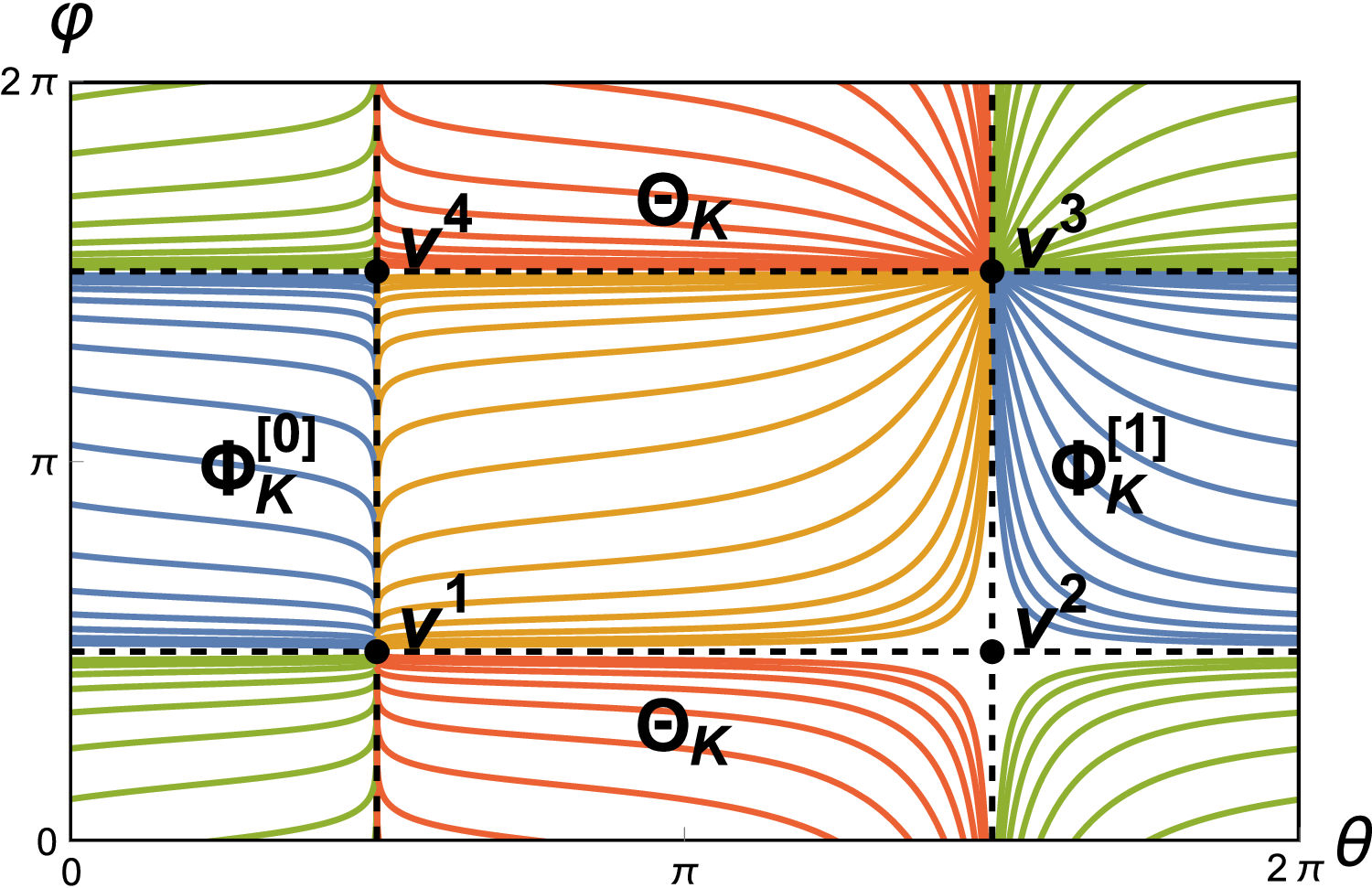}}
\caption{\small Family $\Sigma_K^{(k_1,k_2,0)} (x;\gamma)$ connects minima $v^1$ and $v^3$ on the torus and in the $\theta-\varphi-$plane. Even if only kinks on the front of the torus have been shown in the first figure, the kink variety on the back is identical by symmetries.}\label{4vFam1}
\end{figure}
    
    \item Kink family with $\epsilon=1$: For these solutions minima $v^2$ and $v^4$ are now connected in each region of the torus, see Figure \ref{4vFam2}. In this family the limits for the parameter $\gamma$ are swapped from the previous family. A $\Phi_K^{[1]}-$kink and a $\Theta_K-$kink are obtained for $\gamma\rightarrow -\infty$ while another $\Theta_K-$kink and a $\Phi_K^{[0]}$ are found for $\gamma\rightarrow \infty$. Of course, this is due to the fact that solutions of this family can be obtained applying the previous family of the symmetries that relate the four superpotentials:
\begin{equation}
    (\theta,\varphi)\rightarrow \left(\theta,\varphi+(2l+1)\pi\right) \, , \qquad (\theta,\varphi)\rightarrow \left(\theta,2\pi l-\varphi\right) \, , 
\end{equation}
for $l\in \mathbb{Z}$, which change their topological cluster. Indeed, members $\Sigma_K^{(k_1,k_2,0)} (x;\gamma)$ are related to members $\Sigma_K^{(k_1,k_2+1,1)} (x;-\gamma)$.      
    \begin{figure}[h]
\centerline{\includegraphics[height=4.8cm]{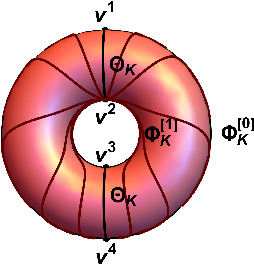}  \hspace{0.5cm} \includegraphics[width=6cm,height=4.5cm]{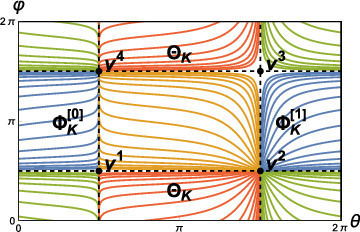}}
\caption{\small Family $\Sigma_K^{(k_1,k_2,1)} (x;\gamma)$ connects minima $v^2$ and $v^4$ on the torus and in the $\theta-\varphi-$plane. Even if only kinks on the front of the torus have been shown in the first figure, the kink variety on the back is identical by symmetries.}\label{4vFam2}
\end{figure}
\end{itemize}
 The limits $\gamma\rightarrow \pm \infty$ of these families are combinations of singular kinks that respect the previously found energy sum rules (\ref{eq:EnergySumRule}). Let us write symbolically those for the two families emerging in the region determined by $k_1=k_2=0$:
\begin{equation}
\lim_{\gamma \rightarrow -\infty} \Sigma_K^{(0,0,0)} (x;\gamma) \equiv \Phi_K^{[0]}(x) \cup \Theta_K (x)\, ,\hspace{0.6cm} \lim_{\gamma \rightarrow \infty} \Sigma_K^{(0,0,0)} (x;\gamma) \equiv \Theta_K (x) \cup \Phi_K^{[1]}(x) \, , \label{familylimit4v}
\end{equation}
\begin{equation}
\lim_{\gamma \rightarrow -\infty} \Sigma_K^{(0,0,1)} (x;\gamma) \equiv \Phi_K^{[1]}(x) \cup \Theta_K (x),\hspace{0.6cm} \lim_{\gamma \rightarrow \infty} \Sigma_K^{(0,0,1)} (x;\gamma) \equiv \Theta_K (x) \cup \, \Phi_K^{[0]}(x) \, . \label{familylimit4vb}
\end{equation}
    This behavior can be observed in Figure \ref{fig:4vDensidadesEnergia}, where in each picture the corresponding energy density profiles with two lumps are recovered when $|\gamma|$ is big enough. Notice that among these pairs of lumps all the three different lumps $\Phi_K^{[0]}$, $\Phi_K^{[1]}$ and $\Theta_K$ can be identified. Indeed, they can be interpreted as three different extended particles.
\begin{figure}[h]
\begin{center}
    \includegraphics[height=3.5cm]{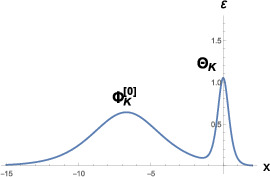} \hspace{0.3cm}
    \includegraphics[height=3.5cm]{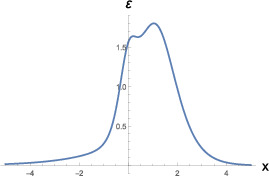} \hspace{0.3cm}
    \includegraphics[height=3.5cm]{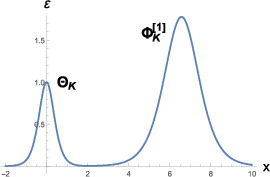}
    \caption{\small Energy densities for three members of the family $\Sigma_K^{(0,0,0)} (x;\gamma)$ given by $\gamma=1$, $\gamma=4.5$ and $\gamma=10$ respectively. Values of parameters $(R,r,m_1,m_2)=(2,0.5,0.5,2)$ have been used. Notice how combinations of singular kinks are obtained as limits. Since orbits of both families $\epsilon=\pm 1$ and in different regions are related by symmetries, these are also the energy densities for the family $\epsilon=1$ and for other regions for certain values of $\gamma$. }\label{fig:4vDensidadesEnergia}
    \end{center}
\end{figure}

In sum, two related families of topological kinks are replicated in the four regions of the torus. Furthermore, the limits of both families coincide with combinations of singular kinks, which delimit these four regions. Since there is no points in these kinks where the flow is undefined, there are no conjugate points where the kink orbits intersect and all these kinks are stable, see \cite{Gonzalez}.

\end{itemize}

\subsection{Kink variety for a model with two vacua (Case 2)}
Models with two vacua can be obtained when $|4 n_1 n_2|=2$. For this condition only two possibilities are available, $n_1=1$ , $n_2=\frac{1}{2}$ and $n_1=\frac{1}{2}$ , $n_2=1$. These configurations correspond to Case $2$ and Case $3$ respectively. Let us explore the first choice of $n_i$, which produces a potential function of the form
\begin{equation}
V_{1,\frac{1}{2}}(\theta,\varphi)= \frac{1}{2} \left[\frac{m_1^2}{r^2} \, \cos^2\theta +\frac{m_2^2}{4(R+r \sin \theta)^2} \, \cos^2 \frac{\varphi}{2}\right]  \, , \label{potentialA2}
\end{equation}
for which the set of vacua ${\cal M}$ is distributed in one side of the torus
\begin{eqnarray*}
{\cal M}_{1,\frac{1}{2}} & = & \textstyle \{ v^1=(\frac{\pi}{2},\pi); v^2=(\frac{3\pi}{2},\pi)\} \, ,
\end{eqnarray*} 
see Figure \ref{fig:ModBsing}. Unlike the previous model, the four superpotentials are not periodic in the torus
\begin{equation}
W_{1,\frac{1}{2}}(\theta,\varphi)=(-1)^{\epsilon_1} \, m_1 \, \sin \theta + (-1)^{\epsilon_2} \, m_2 \, \sin \frac{\varphi}{2} \, .  \label{superpotentialA}
\end{equation}
 This non-periodicity of the superpotential in the torus is precisely what allows the existence of BPS non-topological kinks, just as in \cite{A}. Indeed, if it were periodic, the energy (\ref{eq:EnergyBPS}) of BPS non-topological kinks would vanish, and it could not describe a kink. As expected, reducing the number of vacua of the potential (\ref{potentialA2}) also affects the symmetries between solutions of the superpotential. Only translations $\theta \to \theta+\pi$ relate solutions from different superpotentials now. This splits the torus into two regions where the kink variety has the same structure. In particular, the kink variety in each region includes four singular topological kinks and antikinks and two families of topological kinks:

\begin{itemize}
	\item Singular $\Phi$-kinks: In this case depending on $k_1=0,1$ these singular kinks will cross exclusively $v^1$ or $v^2$
\begin{equation}
\Phi_K^{(k_1,k_2)} (x)= \left(\frac{2k_1+1}{2} \pi \,\, , \,\, 2 \pi (k_2+1)+ 2 {\rm Gd}\left[ (-1)^{\epsilon_2+1} \frac{m_2 \bar{x}}{4(R+(-1)^{k_1}r)^2}\right] \, \right) \, .
\end{equation}
that is, these are non-topological kinks. Here $k_2$ labels solutions under and above $\varphi=\pi$ and $k_1$ will distinguish between those kinks in the ``exterior'' and ``interior'' circumferences. Two types of $\Phi_K-$kinks are identified depending on $k_1$, each one with different energy density profile 
    \begin{equation}
        \varepsilon\left(\Phi_K(x)\right)=\frac{m_2^2}{4 \left(R + (-1)^{k_1} r \right)^2} ~ {\rm sech}^2\left(\frac{m_2 \bar{x}}{4 \left(R +  (-1)^{k_1} r \right)^2}\right) \, ,
    \end{equation}
while the energy remains the same $E[\Phi_K]= 2m_2$ for all values of $k_1$. Same notation as in the previous section $\Phi_K^{[k_1]}$ will be used to distinguish between different energy density profiles. The $\Phi_K^{[0]}-$kink, whose energy density peak is more pronounced, corresponds to the ``exterior'' kink while the $\Phi_K^{[1]}-$kink corresponds to the ``interior'' one, see Figure \ref{fig:ModBsing}. Therefore, these singular kinks describe two different vrochosons, because they cannot decay to a vacuum solution.
\begin{figure}[ht]
\begin{center}
    \includegraphics[height=4.5cm]{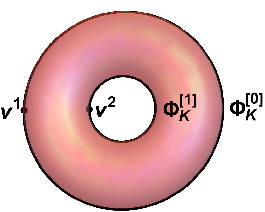} \hspace{0.5cm}
    \includegraphics[height=4.5cm]{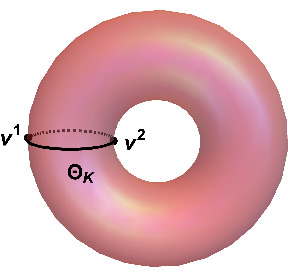}
    \caption{\small $\Phi_K$ and $\Theta_K-$kink orbits are represented on the torus, where dotted lines show kinks on the back side. See how these singular kinks combined split the torus into two regions related by symmetries of the potential.}\label{fig:ModBsing}
    \end{center}
\end{figure}

\item Singular $\Theta$-kinks: $\varphi$-constant orbits produce in the first order equations (\ref{eq:PrimerOrden}) two topological kinks and antikinks 
\begin{equation}
\Theta_K^{(k_1,k_2)} (x) =  \left((k_1+1) \pi+ {\rm Gd}\left[ (-1)^{k_1+\epsilon_1+1} \frac{m_1 \bar{x}}{r^2}\right]\, ,  \pi \right) \hspace{0.3cm},\hspace{0.3cm} \quad k_1=0,1 \, ,
\end{equation}
which correspond to half-circumferences associated to the minor radius, see Figure \ref{figure:A1y2}. These poloidal curves will connect asymptotically the two minima $v^1$ and $v^2$. The two values of $[k_1]$ distinguish between the kink on the front and the kink on the back of the torus, see Figure \ref{fig:ModBsing}. Once again, both the energy density and the energy of all these $\Theta_K(x)$-kinks are identical:
\[
\varepsilon\left(\Theta_K(x)\right)=\frac{m_1^2}{r^2} ~ {\rm sech}^2\left(\frac{m_1 \bar{x}}{r^2}\right) \, , \qquad E[\Theta_K]=2 m_1 \, .
\]
\item Kink families: The value $n_1$ has not changed and therefore the same reparametrization $\xi(x)$ as in Case $1$ is obtained. Alternatively, the explicit orbit equation is shown, which provides the family of kinks
\begin{equation}
\sin \frac{\varphi}{2}=  \tanh \left[\gamma + (-1)^{\epsilon} \, \frac{m_2}{8 m_1} \frac{r^2}{(R+r)^2} \ln \left|\frac{(1+ \sin \theta)^{\left(\frac{R+r}{R-r}\right)^2} e^{\left(\frac{R+r}{R-r} \frac{2r}{R+r\sin \theta}\right)}}{(1-\sin \theta)(R+r \sin \theta)^{\frac{4 r R}{(R-r)^2}}}\right|\right] \, ,
\label{eq:orbB}
\end{equation}
where the parameter $\gamma\in (-\infty,\infty )$ is an integration constant, see Figure \ref{2vFam1}. Notice that in this case both values $\epsilon=0,1$ describe the same orbits on the torus since they complete a whole revolution along the toroidal direction. Members of these families shall be denoted as $\Sigma_K^{(k_1,k_2,\epsilon)} (x;\gamma)$, where the parameter $\gamma$ is again employed to label every kink member in each family.
\begin{figure}[h]
\centerline{\includegraphics[height=4.8cm]{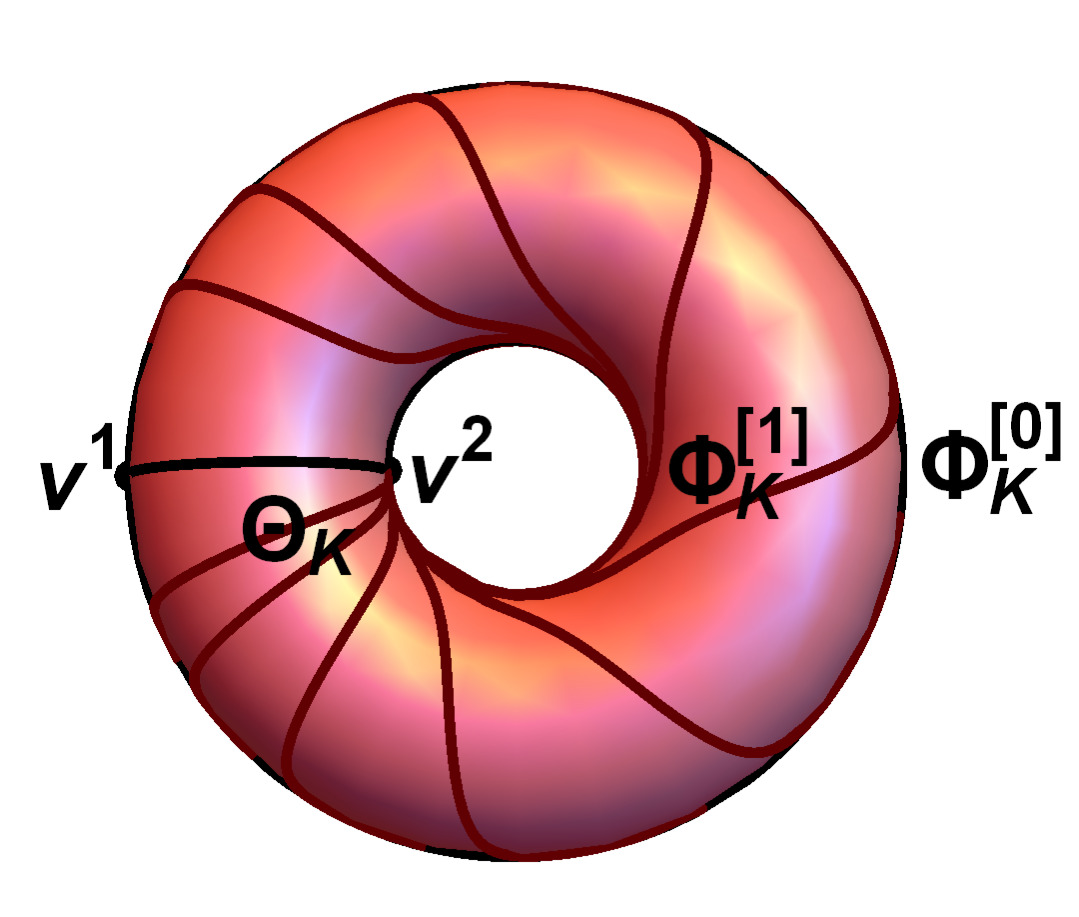}  \hspace{0.5cm} \includegraphics[width=6cm,height=4.5cm]{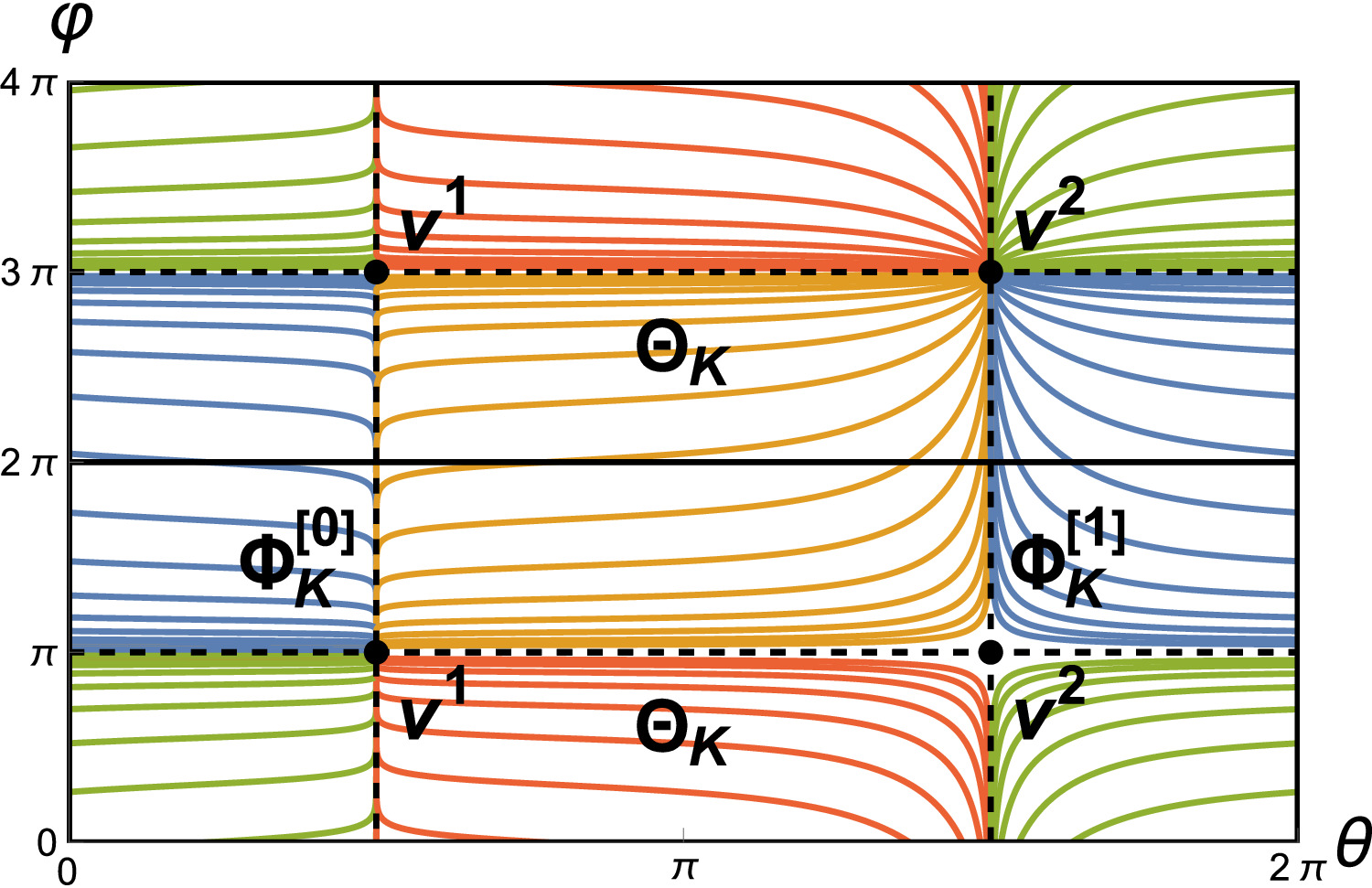}}
\caption{\small Family $\Sigma_K^{(k_1,k_2,0)} (x;\gamma)$ connects minima $v^1$ and $v^2$ on the torus and in the $\theta-\varphi-$plane. Even if only kinks on the front of the torus have been shown in the first figure, the kink variety on the back is identical by symmetries. Given the identification between members of both families, orbits for the family $\Sigma_K^{(k_1,k_2,1)} (x;\gamma)$ coincide with those of $\Sigma_K^{(k_1,k_2+1,0)} (x;\gamma)$.}\label{2vFam1}
\end{figure}
In this case both families, $\epsilon=0$ and $\epsilon=1$, connect the same minima $v^1$ and $v^2$ in both regions of the torus. In fact, members of both families describe the same kink orbits and therefore only one family is found in each region. Indeed, a symmetry between superpotentials:
\begin{equation}
    (\theta,\varphi)\rightarrow \left(\theta,\varphi+(2l+1) 2\pi\right) \, , \qquad (\theta,\varphi)\rightarrow \left(\theta,4\pi l-\varphi\right) \, , 
\end{equation}
for $l\in \mathbb{Z}$, allows us to make an identification between members $\Sigma_K^{(k_1,k_2,0)} (x;\gamma)=\Sigma_K^{(k_1,k_2+1,1)} (x;-\gamma)$ of both families, which in this case belong to the same topological cluster. Two copies of the same torus have been represented in the $\theta-\varphi-$plane so that both related solutions can be seen in the $\theta-\varphi-$plane. On the other hand, limits of the orbit equation for $\gamma\rightarrow \pm \infty$ are almost identical to those of the previous model. Let us restrict to the region $k_1=k_2=0$. The obtained limit values for $\varphi$ have been doubled, that is $\varphi=\pi+4 \pi l_1$ and $\varphi=3\pi+4 \pi l_1$, and so has been its periodicity in this direction. These of course correspond to the analogue $\Theta-$kinks in this model. In sum, for the first family $\epsilon=0$ a $\Phi^{[1]}_K-$kink and a $\Theta_K-$kink are recovered for $\gamma\rightarrow -\infty$ while another $\Theta_K-$kink and a $\Phi^{[0]}_K-$kink for $\gamma\rightarrow \infty$ are obtained. Hence, these families have as limits the same combinations of singular kinks as in the previous case, which obviously also abide by the energy sum rules. These combinations are symbolically written the same way as before 
\begin{equation}
\lim_{\gamma \rightarrow -\infty} \Sigma_K^{(0,0,0)} (x;\gamma) \equiv \Phi_K^{[0]}(x) \cup \Theta_K (x)\, ,\hspace{0.6cm} \lim_{\gamma \rightarrow \infty} \Sigma_K^{(0,0,0)} (x;\gamma) \equiv \Theta_K (x) \cup \Phi_K^{[1]}(x) \, . \label{familylimit21}
\end{equation}
\begin{equation}
\lim_{\gamma \rightarrow -\infty} \Sigma_K^{(0,0,1)} (x;\gamma) \equiv \Phi_K^{[1]}(x) \cup \Theta_K (x)\, ,\hspace{0.6cm} \lim_{\gamma \rightarrow \infty} \Sigma_K^{(0,0,1)} (x;\gamma) \equiv \Theta_K (x) \cup \Phi_K^{[0]}(x) \, . \label{familylimit22}
\end{equation}
 and can be visualized in Figure \ref{fig:2v1hDensidadesEnergia}, where two extended particles can be identified for each limit of $\gamma$. Lastly, same analysis as in the previous model reveals that, due to the absence of intermediate conjugate points, these families of kinks are stable.
 \begin{figure}[h]
\begin{center}
    \includegraphics[height=3.5cm]{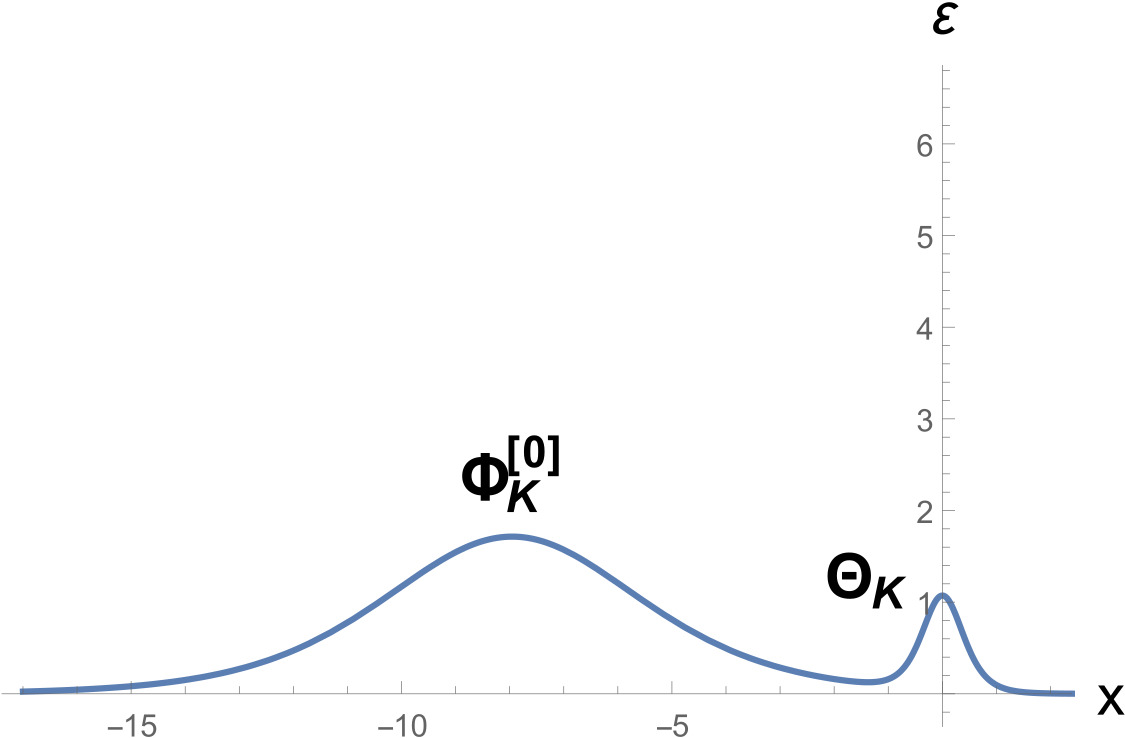} \hspace{0.3cm}
    \includegraphics[height=3.5cm]{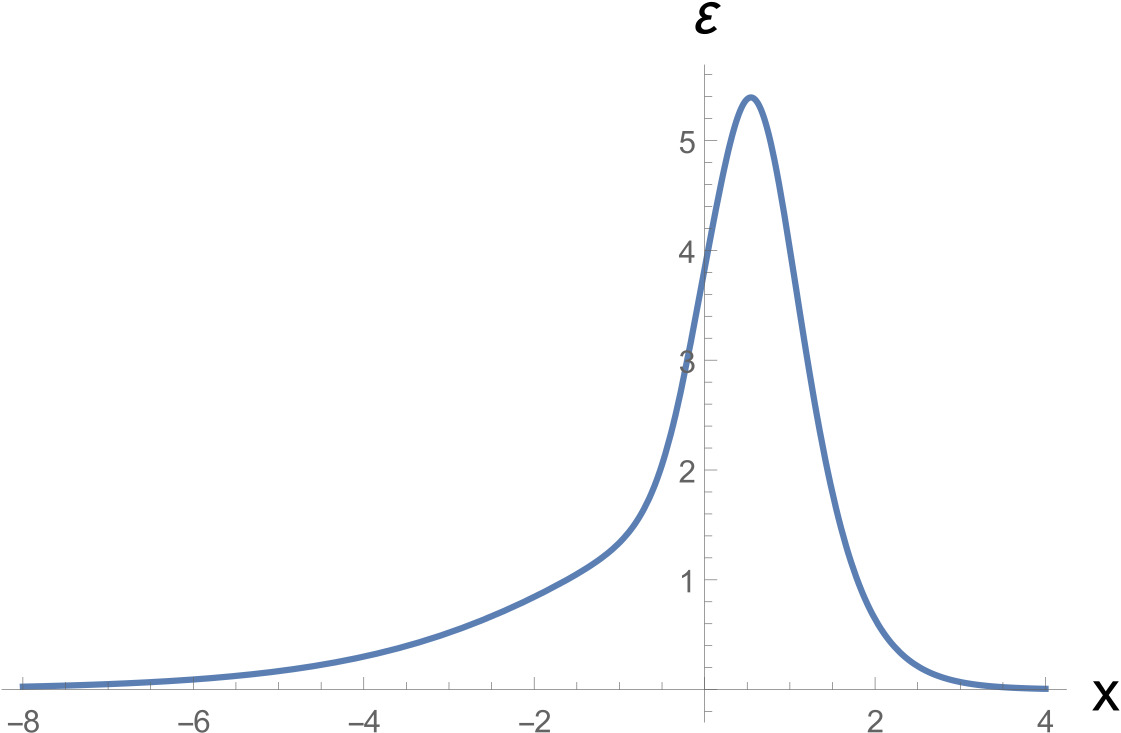} \hspace{0.3cm}
    \includegraphics[height=3.5cm]{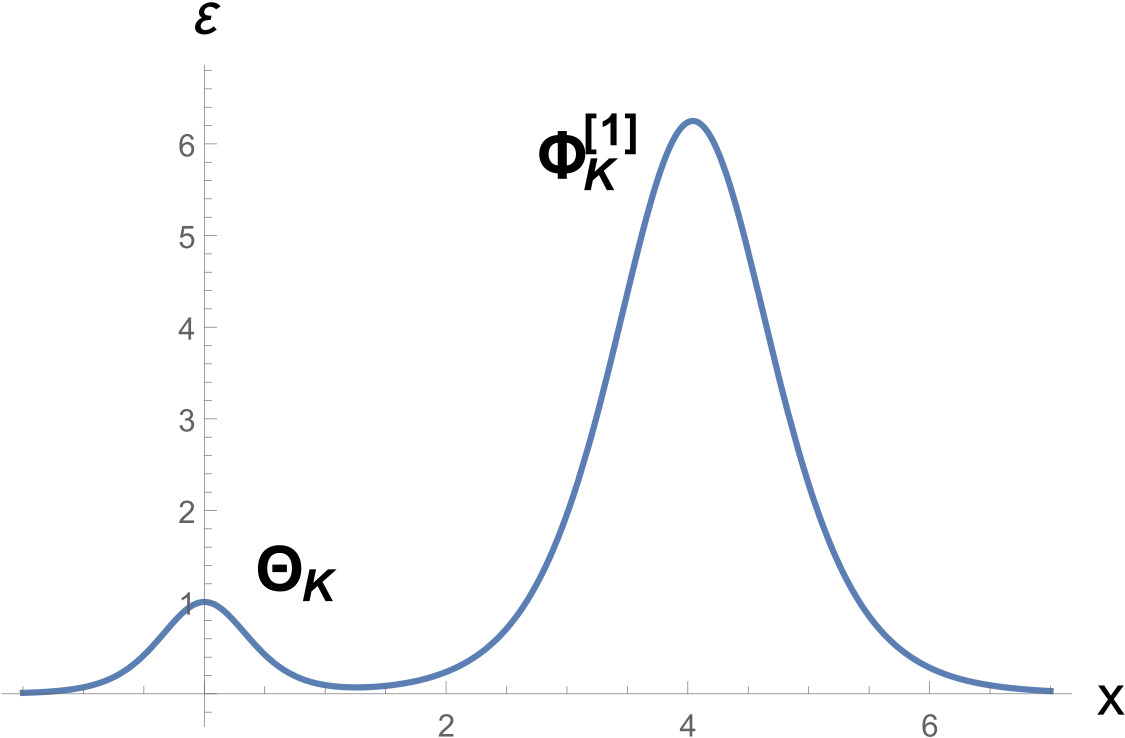}
    \caption{\small Energy densities for three members of the family $\Sigma_K^{(0,0,0)} (x;\gamma)$ given by $\gamma=-3$, $\gamma=1$ and $\gamma=5$ respectively. Values of parameters $(R,r,m_1,m_2)=(1.6,0.5,0.5,5.5)$ have been used. Notice how combinations of singular kinks are obtained as limits. Since orbits of both families $\epsilon=\pm 1$ and in different regions are related by symmetries, these are also the energy densities for the family $\epsilon=1$ and for other regions for certain values of $\gamma$. }\label{fig:2v1hDensidadesEnergia}
    \end{center}
\end{figure}

\end{itemize}

\subsection{Kink variety for a model with two vacua (Case 3)}
The remaining configuration of values of $n_1$ and $n_2$ that produces a potential (\ref{potential}) with two minima is $n_1=\frac{1}{2}$ and $n_2=1$. Even if the number of minima is the same as in Case $2$, their distribution on the torus and the kink variety will be different. This corresponds to Case $3$, which will be now studied. 
In particular, this choice of $n_i$ produces a potential
\begin{equation}
V_{\frac{1}{2},1}(\theta,\varphi)= \frac{1}{2} \left[\frac{m_1^2}{4r^2} \, \cos^2 \frac{\theta}{2} +\frac{m_2^2}{(R+r \sin \theta)^2} \, \cos^2 \varphi\right] \, , \label{potentialB}
\end{equation}
which can be derived from the following four superpotentials, one for each value combination of $\epsilon_i=\pm 1$
\begin{equation}
W_{\frac{1}{2},1}(\theta,\varphi)=(-1)^{\epsilon_1} \, m_1 \, \sin \frac{\theta}{2} + (-1)^{\epsilon_2} \, m_2 \, \sin \varphi \, .   \label{superpotentialB}
\end{equation}
 Once again, the fact that these superpotentials are not periodic on the torus allows the existence of BPS non-topological kinks. However, since the two minima are aligned in the poloidal direction
\[
{\cal M}_{\frac{1}{2},1}=\left\{ v^1= \left(\pi,\frac{\pi}{2} \right)\,\, , \,\, v^2= \left(\pi,\frac{3\pi}{2}\right) \right\} \, ,
\]
as it is shown in Figure \ref{fig:2vh1Singulares}, the singular non-topological kinks in this model will be poloidal and not toroidal. On the other hand, in this case only translations $\varphi \to \varphi+\pi$ will identify solutions from different superpotentials. This splits the torus into two equivalent regions again. Notice, however, that these regions are different from those of Case $2$. In each of these two regions two singular topological kinks, two singular non-topological kinks and two families of non-topological kinks are found:

\begin{itemize}
	\item Singular $\Phi$-kinks: General expression (\ref{eq:SingularPhi}) for toroidal orbits provide now topological singular kinks because of the new alignment of the minima:
\begin{equation}
\Phi_K^{(k_1,k_2)}(x)= \left((2k_1+1)\pi \,\, , \,\, (k_2+1)\pi + ~ {\rm Gd}\left[ (-1)^{k_2+\epsilon_2+1} \, \frac{m_2 \overline{x}}{R^2}\right]\right) \, .
\end{equation}
Indeed, this expression describes singular topological kinks and their antikinks, which asymptotically connect minima $v^1$ and $v^2$, see Figure \ref{fig:2vh1Singulares}. Notice that depending on $k_1$ and $k_2$ the expression above describes different singular kinks in the $\theta-\varphi-$plane, but on the torus only $[k_2]$ distinguishes between different kinks. In fact, the multiplicity of energy density profiles for $\Phi_K-$kinks of previous models disappears since only one profile emerges
    \begin{equation}
        \varepsilon\left(\Phi_K(x)\right)=\frac{m_2^2}{R^2} ~ {\rm sech}^2\left(\frac{m_2 \bar{x}}{R^2}\right) \, , \qquad E[\Phi_K]=2m_2 \, .
    \end{equation}
and this solution describes only one type of extended particle.
\begin{figure}[ht]
\begin{center}
    \includegraphics[height=4.5cm]{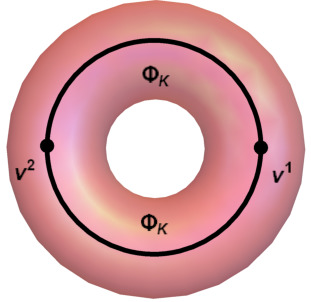} \hspace{0.5cm}
    \includegraphics[height=4.5cm]{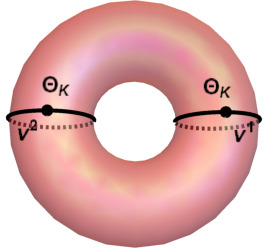}
    \caption{\small $\Phi_K$ and $\Theta_K-$kink orbits are represented on the torus, where dotted lines show kinks on the back side. See how these singular kinks combined split the torus into two regions related by symmetries of the potential.}\label{fig:2vh1Singulares}
    \end{center}
\end{figure}

\item Singular $\Theta$-kinks: Poloidal singular solutions (\ref{eq:SingularTheta}) now read
\[
\Theta_K^{(k_1k_2)} (x)= \left( 2 \,(k_1+1) \pi + 2 ~ {\rm Gd}\left[(-1)^{k_1+k_2+1} \frac{m\overline{x}}{2r}\right]  \,\, , \,\,  \frac{(2k_2+1)}{2} \pi \right) \, , 
\]
and describe a couple of non-topological kinks and their antikinks, that is, two poloidal loops that cross $v^1$ and $v^2$ respectively depending on the value of $k_2$, see Figure \ref{fig:2vh1Singulares}. While $k_1$ labels different kinks in the $\theta-\varphi-$plane, on the torus will describe the same kink. As stated before, the energy density and total energy of these kinks is independent of $k_i$
\[
\varepsilon\left(\Theta_K(x)\right)=\frac{m_1^2}{4r^2} ~ {\rm sech}^2\left(\frac{m_1 \bar{x}}{4r^2}\right) \, \qquad E[\Theta_K]=2 m_1\, ,
\]
describing a second type of vrochoson.

\item Kink families: The new value of $n_1$ makes the expression for the new parameter $\xi(x)$ more complicated, but it is still analytically available. Instead, as in the previous cases, the two families $\epsilon=0,1$ on the torus are displayed by presenting the orbit equation 
\begin{equation}
\sin \varphi=  \tanh \left[\gamma + (-1)^{\epsilon} \, \frac{2 m_2 r^2}{m_1} f(\theta)\right] \, ,
\label{eq:orb3}
\end{equation}
where the parameter $\gamma\in (-\infty,\infty )$ is an integration constant and the function $f$ is defined as follows 
\[
    f(\theta)=\frac{2 \log \left(\left| 1+\frac{2}{\cot \left(\frac{\theta }{4}\right)-1}\right| \right)}{R^2}-\frac{2 r \left(r \sin \left(\frac{\theta }{2}\right)+R \cos
   \left(\frac{\theta }{2}\right)\right)}{R \left(r^2-R^2\right) (r \sin (\theta )+R)}+
\]
\[
   +\frac{(2 r-3 R) \tan ^{-1}\left(\frac{2 r \left(\tan \left(\frac{\theta
   }{4}\right)+1\right) \sqrt{\frac{R}{r}-1}}{2 r \left(\tan \left(\frac{\theta }{4}\right)+1\right)-R \sec ^2\left(\frac{\theta }{4}\right)}\right)}{r R^2
   \left(\frac{R}{r}-1\right)^{3/2}}+\frac{\sqrt{r} (2 r+3 R) \tanh ^{-1}\left(\frac{\sqrt{r} \left(\cos \left(\frac{\theta }{2}\right)-\sin \left(\frac{\theta
   }{2}\right)\right)}{\sqrt{r+R}}\right)}{R^2 (r+R)^{3/2}} \, .
\]
  Solutions of this orbit equation connect $v^1$ and $v^2$, see Figure \ref{2vh1Fam1}. Exactly as before, members of these solutions shall be denoted as $\Sigma_K^{(k_1,k_2,\epsilon)} (x;\gamma)$, where each family member is labeled by the parameter $\gamma$. In this case, once more, neither of the following symmetries between superpotentials:
\begin{equation}
    (\theta,\varphi)\rightarrow \left(\theta,\varphi+(2l+1)\pi\right) \, , \qquad (\theta,\varphi)\rightarrow \left(\theta,2\pi l-\varphi\right) \, , 
\end{equation}
for $l\in \mathbb{Z}$, change the topological sector of the solutions. Similarly to previous cases, these transformations generate a change in the orbit equation which can be absorbed by $\epsilon\rightarrow\epsilon+1$ and $\bar{\gamma}=-\gamma$:
\begin{equation}
\sin \varphi=  \tanh \left[\bar{\gamma} + (-1)^{\epsilon+1} \, \frac{2 m_2 r^2}{m_1} f(\theta)\right] \, .
\end{equation}
This orbit equation corresponds to members of the other family $\epsilon+1$ labeled with parameter $\bar{\gamma}$, which identify members $\Sigma_K^{(k_1,k_2,0)} (x;\gamma)=\Sigma_K^{(k_1,k_2+1,1)} (x;-\gamma)$. This implies that also in this case only one family exists in each region of the torus. Again, two copies of the same torus have been represented in the $\theta-\varphi-$plane so that both related solutions can be seen.
    \begin{figure}[h]
\centerline{\includegraphics[height=4.5cm]{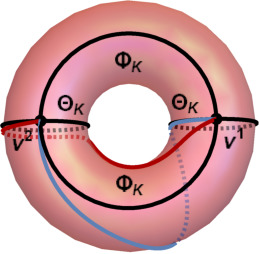} \hspace{0.5cm}
\includegraphics[width=6cm,height=4.5cm]{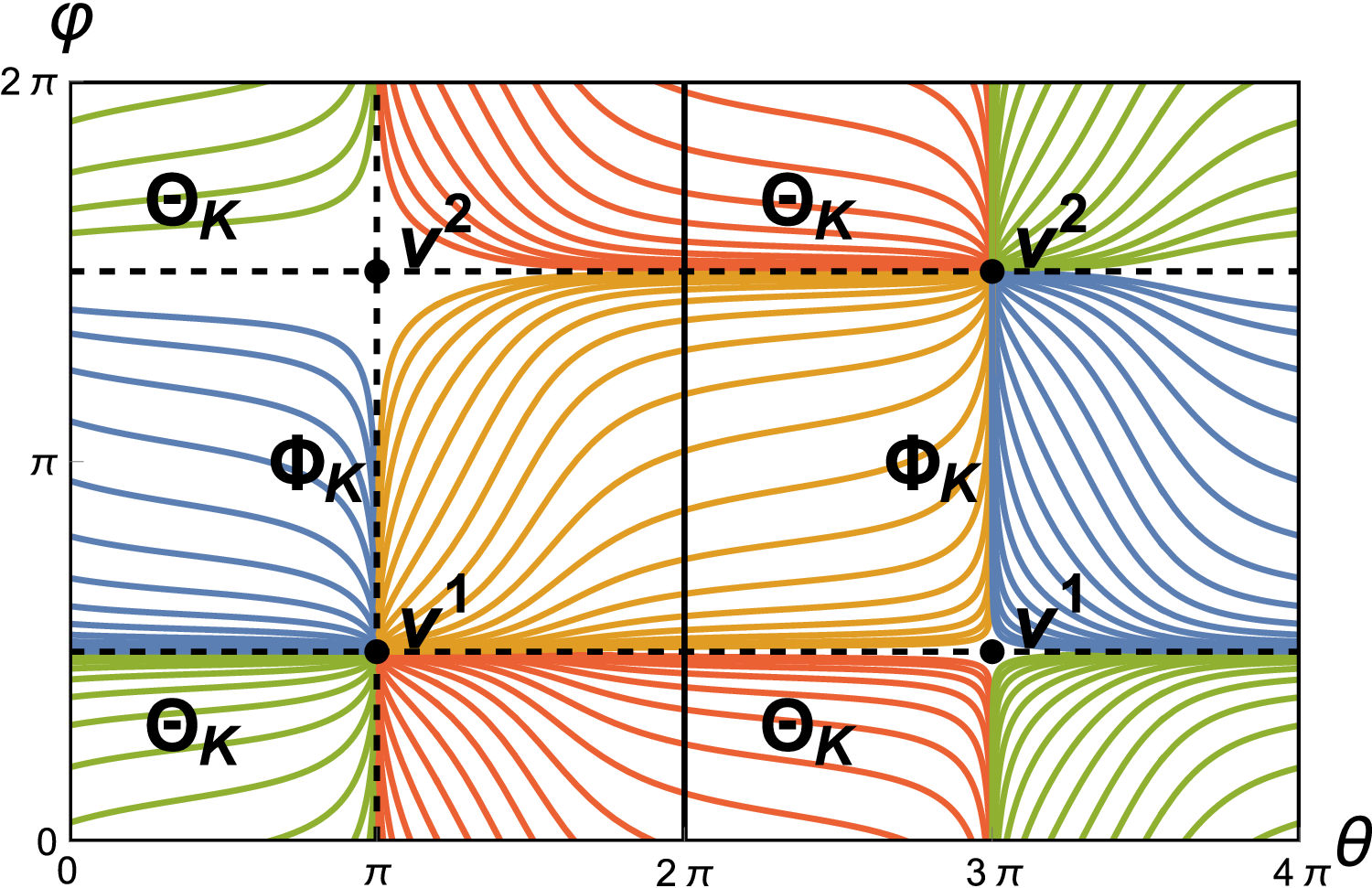}}
\caption{\small Family $\Sigma_K^{(k_1,k_2,0)} (x;\gamma)$ connects minima $v^1$ and $v^2$ on the torus and in the $\theta-\varphi-$plane. The two solutions depicted on the torus correspond to two solutions $\Sigma_K^{(0,0,0)} (x;\gamma)$ in the $\theta-\varphi-$plane for $\gamma=-2$ (red) and $\gamma=2$ (blue).} \label{2vh1Fam1}
\end{figure}   
Limits of the orbit equation are similar to those of previous models. Considering that $\displaystyle\lim_{\theta\rightarrow \pm\pi}f(\theta)=\pm\infty$, it is easy to prove that a $\Phi_K-$kink and a $\Theta_K-$kink on one hand and a $\Theta_K$ and a $\Phi_K-$kink are recovered as limits for the family $\epsilon=0$ for $\gamma\rightarrow -\infty$ and $\gamma\rightarrow \infty$ respectively in the region labeled by $k_1=k_2=0$. Symmetries relating these families immediately provide the limits for the other family, which only swaps the positions of the left and right $\Theta-$kink. All these kinks that belong to these two families are topological kinks that connect $v^1$ to $v^2$. Therefore, the transformations between families leave the topological sector invariant again. As would be expected, only two types of extended particles appear in the limits of these families
\begin{equation}
\lim_{\gamma \rightarrow -\infty} \Sigma_K^{(0,0,0)} (x;\gamma) \equiv \Phi_K(x) \cup \Theta_K (x)\, ,\hspace{0.6cm} \lim_{\gamma \rightarrow \infty} \Sigma_K^{(0,0,0)} (x;\gamma) \equiv \Theta_K (x) \cup \Phi_K(x) \, . \label{familylimit3}
\end{equation}
\begin{equation}
\lim_{\gamma \rightarrow -\infty} \Sigma_K^{(0,0,1)} (x;\gamma) \equiv \Phi_K(x) \cup \Theta_K (x)\, ,\hspace{0.6cm} \lim_{\gamma \rightarrow \infty} \Sigma_K^{(0,0,1)} (x;\gamma) \equiv \Theta_K (x) \cup \Phi_K(x) \, . \label{familylimit3b}
\end{equation}
which can be observed in Figure \ref{fig:2vh1DensidadesEnergia}. Finally, no intermediate conjugate points along the kink orbits gives us another proof of their stability. 

\end{itemize}

\begin{figure}[h]
\begin{center}
    \includegraphics[height=3.5cm]{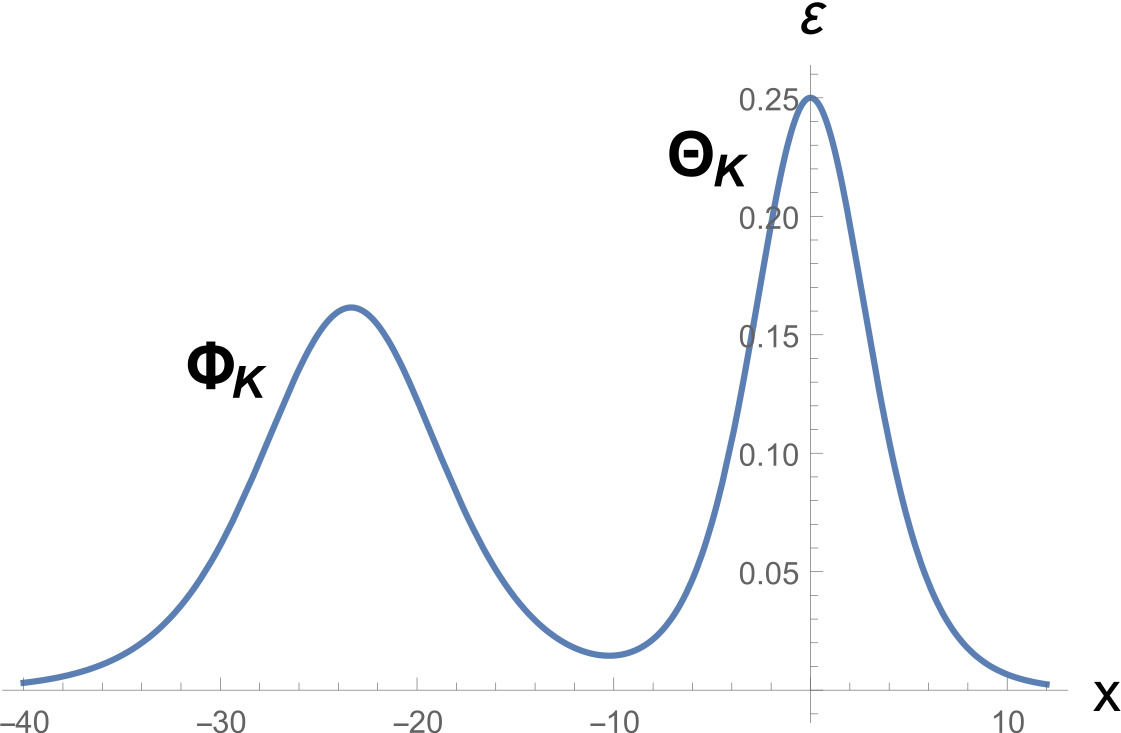} \hspace{0.3cm}
    \includegraphics[height=3.5cm]{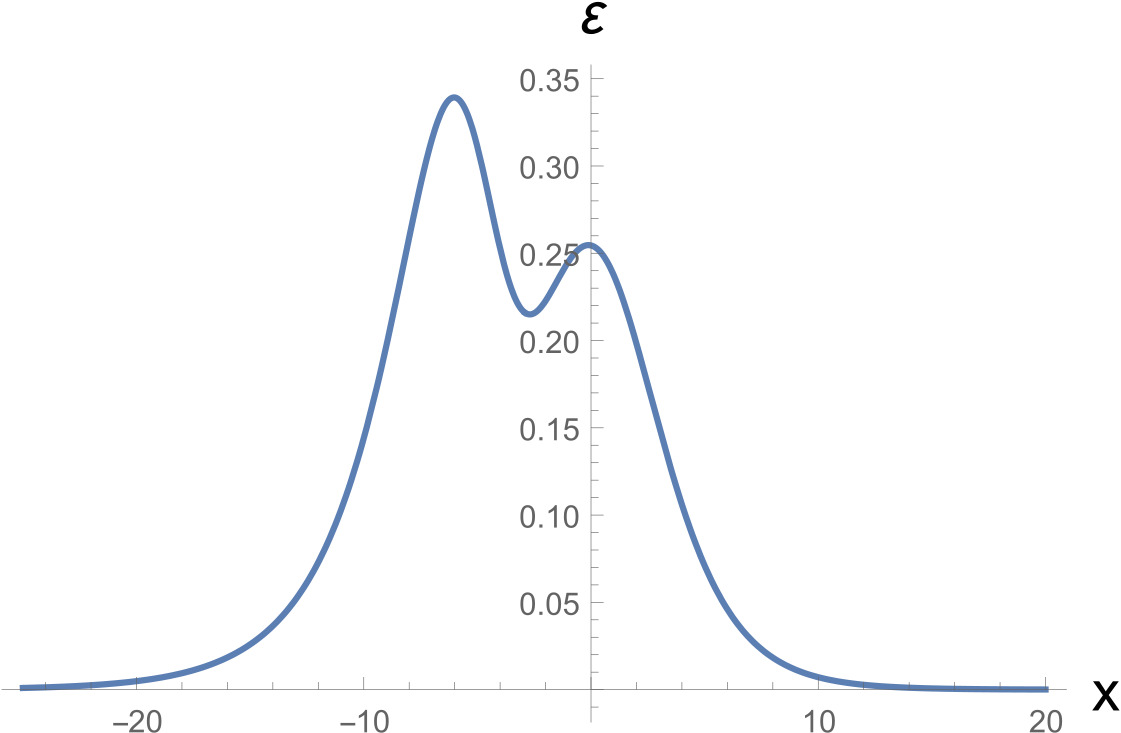} \hspace{0.3cm}
    \includegraphics[height=3.5cm]{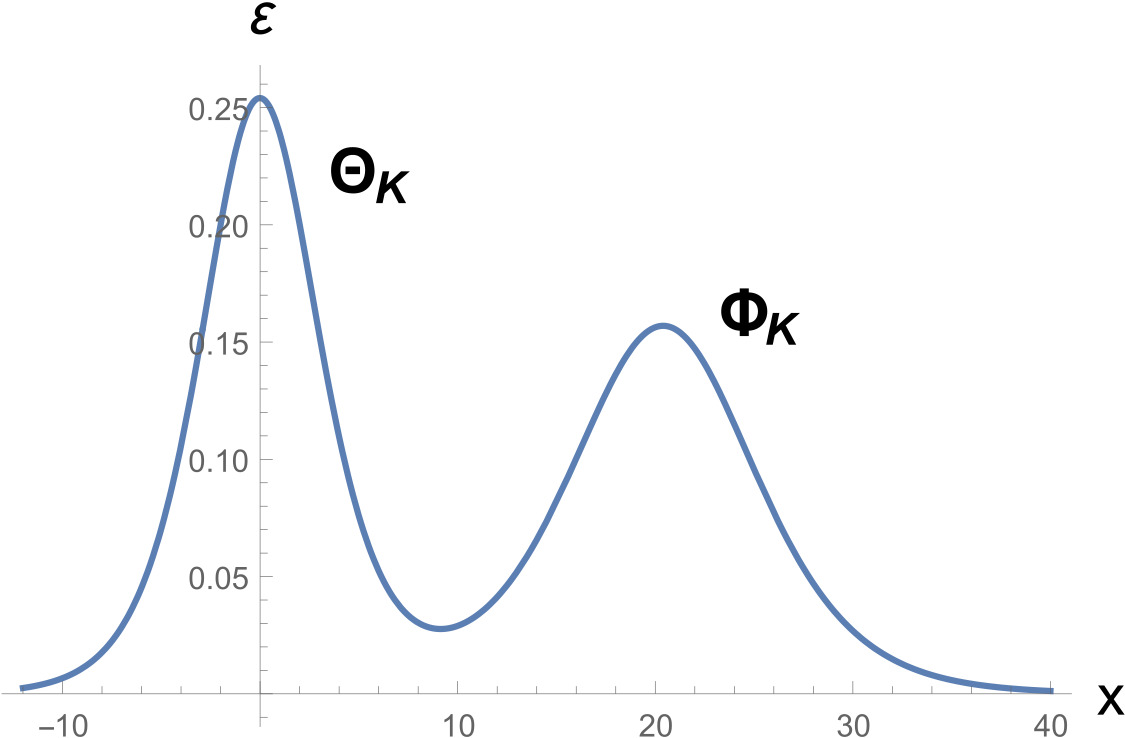}
    \caption{\small Energy densities for three members of the family $\Sigma_K^{(0,0,0)} (x;\gamma)$ given by $\gamma=-3$, $\gamma=0$ and $\gamma=5$ respectively. Values of parameters $(R,r,m_1,m_2)=(2.5,1,1,1)$ have been used. Notice how combinations of singular kinks are obtained as limits. Since orbits of both families $\epsilon=\pm 1$ and in different regions are related by symmetries, these are also the energy densities for the family $\epsilon=1$ and for other regions for certain values of $\gamma$. }\label{fig:2vh1DensidadesEnergia}
    \end{center}
\end{figure}

\subsection{Kink variety for a model with one vacuum point}
 The number of vacua of our models $|4 n_1 n_2|$ is minimized when the values $n_1=n_2=\frac{1}{2}$ are taken to the potential (\ref{potential}) and is equal to $1$. This is precisely the scenario in Case $4$, where this choice of $n_i$ leads to a model with only one minimum. In particular, these conditions produce a potential of the form
\begin{equation}
V_{\frac{1}{2},\frac{1}{2}}(\theta,\varphi)= \frac{1}{8} \left[\frac{m_1^2}{r^2} \, \cos^2 \frac{\theta}{2} +\frac{m_2^2}{(R+r \sin \theta)^2} \, \cos^2 \frac{\varphi}{2}\right] \, , \label{potentialB}
\end{equation}
which can be derived in this last case from these four superpotentials
\begin{equation}
W_{\frac{1}{2},\frac{1}{2}}(\theta,\varphi)=(-1)^{\epsilon_1} \, m_1 \, \sin \frac{\theta}{2} + (-1)^{\epsilon_2} \, m_2 \, \sin \frac{\varphi}{2} \, .   \label{superpotentialB}
\end{equation}
 Observe that in this model the superpotentials are non-periodic on the torus in both angles. This fact once again allows the existence of BPS non-topological kinks, but in both directions in this case. As intended, the set ${\cal M}$ contains now only a minimum
\[
{\cal M}_{\frac{1}{2},\frac{1}{2}}=\left\{ v^1= \left(\pi,\pi \right) \right\} \, ,
\]
as it can be observed in Figure \ref{fig:1vSingulares}. Notice that every symmetry that related solutions from different superpotentials on the torus disappear given its form. The kink variety in the torus will include two singular non-topological kinks and a family of non-topological kinks. Indeed, the existence of only one vacuum point forces every kink to be non-topological. A thorough description of these kinks is as follows:

\begin{itemize}
	\item Singular $\Phi$-kinks: Naturally, since there is only one minimum on the torus no more than one kink of this type can be found 
\begin{equation}
\Phi_K^{(k_1,k_2)}(x)= \left((2k_1+1)\pi \,\, , \,\, 2(k_2+1) \pi+  2 ~ {\rm Gd}\left[ (-1)^{\epsilon_2+1} \, \frac{m_2 \overline{x}}{4R^2}\right]\right) \, ,
\end{equation}
see Figure \ref{fig:1vSingulares}. Once again, singular kinks with different $k_i$ in the $\theta-\varphi-$plane will represent the same kink on the torus. Its energy density and total energy are of the form 
\[
        \varepsilon\left(\Phi_K(x)\right)=\frac{m_2^2}{4 R^2} ~ {\rm sech}^2\left(\frac{m_2 \bar{x}}{4 R^2}\right) \, , \qquad E[\Phi_K]=2m_2 \, .
\]
\begin{figure}[ht]
\begin{center}
    \includegraphics[height=4.5cm]{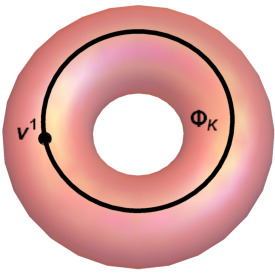} \hspace{0.5cm}
    \includegraphics[height=4.5cm]{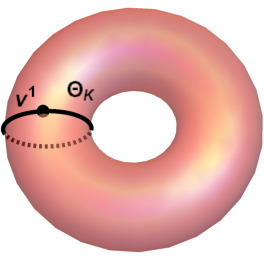}
    \caption{\small $\Phi_K$ and $\Theta_K-$kink orbits are represented on the torus, where dotted lines represent the part of the $\Theta_K-$kink orbit on the back side.}\label{fig:1vSingulares}
    \end{center}
\end{figure}

\item Singular $\Theta$-kinks: For the same reason as in the $\Phi-$kinks, only one type of $\Theta-$kink can emerge in this model
\[
\Theta_K^{(k_1, k_2)} (x)= \left( 2 \,(k_1+1) ~ \pi + 2 ~ {\rm Gd}\left[  (-1)^{k_1+\epsilon_1+1} \frac{m_1\overline{x}}{4r^2}\right]  \,\, , \,\,  (2k_2+1)\pi \right) \, , 
\]
 where $k_1$ and $k_2$ will affect the position in the $\theta-\varphi-$plane but not on the torus, see Figure \ref{fig:1vSingulares}. The energy density and total energy of these kinks are
\[
\varepsilon\left(\Theta_K(x)\right)=\frac{m_1^2}{4r^2} ~ {\rm sech}^2\left(\frac{m_1 \bar{x}}{4r^2}\right) \, , \qquad E[\Theta_K]=2 m_1\, .
\]
Notice that in this case only one $\Theta$-kink and only one $\Phi$-kink are found.
\item Kink families: The reparametrization $\xi(x)$ is identical to that of Case $3$, but again the orbit equation will be presented instead
\begin{equation}
\sin \frac{\varphi}{2}=  \tanh \left[\gamma + (-1)^{\epsilon} \, \frac{m_2 r^2}{2m_1} f(\theta)\right] \, ,
\label{eq:orb4}
\end{equation}
where the parameter $\gamma\in (-\infty,\infty )$ is again an integration constant and $f$ is the function previously defined. Let us denote once more solutions of these families on the torus as $\Sigma_K^{(k_1,k_2,\epsilon)} (x;\gamma)$ with parameter $\gamma$. The relation between families from previous models persists and plotting one of the families in the $\theta-\varphi-$plane is enough to identify all these non-topological kinks. These two families $\epsilon=0,1$ are still related by one of the transformations
\begin{equation}
    (\theta,\varphi)\rightarrow \left(\theta,\varphi+(2l+1)\pi\right) \, , \qquad (\theta,\varphi)\rightarrow \left(\theta,4\pi l-\varphi\right) \, , 
\end{equation}
for $l\in \mathbb{Z}$, producing the same identification of members $\Sigma_K^{(k_1,k_2,0)} (x;\gamma)=\Sigma_K^{(k_1,k_2+1,1)} (x;-\gamma)$. Indeed, these shifts produce a change in the orbit equation that can be absorbed by swapping the family $\epsilon\rightarrow \epsilon+1$ and redefining the parameter $\bar{\gamma}=-\gamma$:
\[\sin \frac{\varphi}{2}=  \tanh \left[\bar{\gamma} + (-1)^{\epsilon+1} \, \frac{m_2 r^2}{m_1} f(\theta)\right] \, .\]
This implies that both values $\epsilon=0,1$ produce the same orbits on the torus since these orbits complete a revolution in the toroidal direction. Hence, this model exhibits only one family of kinks. On the other hand, these transformations obviously cannot change the topological cluster of solutions since there exists only one. By construction only non-topological kinks are allowed to exist in this model, solutions leave the minimum to return to it. However, within this topological cluster three different topological sectors contain kinks. Let us denote as $(\omega_{\theta},\omega_{\varphi})$ the winding numbers around the poloidal and toroidal directions. Singular kinks revolve once around one of the directions, that is,  $\Theta_K-$kink has $(|\omega_{\theta}|,|\omega_{\varphi}|)=(1, 0)$ while  $\Phi_K-$kink has $(|\omega_{\theta}|,|\omega_{\varphi}|)=(0, 1)$. Here the absolute value is necessary given that a kink and its anti-kink will describe the same orbit. However, unlike these singular kinks, members of the family of kinks revolve simultaneously around both directions $(|\omega_{\theta}|,|\omega_{\varphi}|)=(1, 1)$. Therefore, kinks in this family belong to a third topological sector, even if they belong to the same topological cluster.      
    \begin{figure}[h]
\centerline{\includegraphics[height=4.5cm]{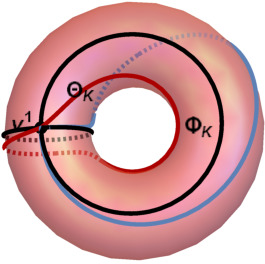} \hspace{0.5cm}\includegraphics[width=6cm,height=4.5cm]{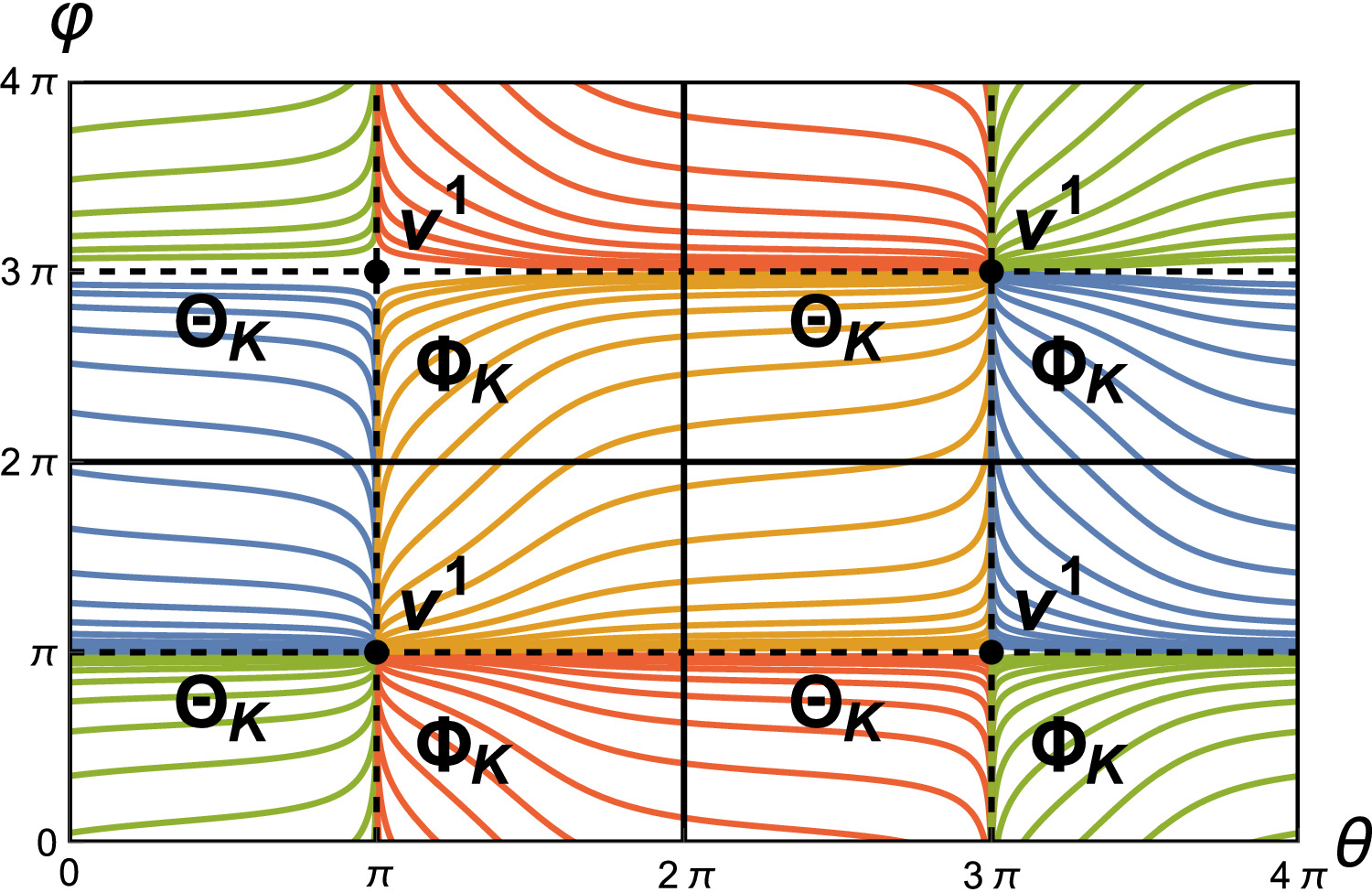}}
\caption{\small Family $\Sigma_K^{(k_1,k_2,0)} (x;\gamma)$ connects minimum $v^1$ with itself on the torus and in the $\theta-\varphi-$plane. The two solutions depicted on the torus correspond to two solutions $\Sigma_K^{(0,0,0)} (x;\gamma)$ in the $\theta-\varphi-$plane for $\gamma=-2$ (red) and $\gamma=2$ (blue).}\label{1vFam1}
\end{figure}

  Identical procedure as in previous sections provides formally the same limits for both families as in the previous model, that is, a combination of a $\Theta_K-$kink and a $\Phi_K-$kink. In this case all these are non-topological 
\begin{equation}
\lim_{\gamma \rightarrow -\infty} \Sigma_K^{(0,0,0)} (x;\gamma) \equiv \Phi_K(x) \cup \Theta_K (x)\, ,\hspace{0.6cm} \lim_{\gamma \rightarrow \infty} \Sigma_K^{(0,0,0)} (x;\gamma) \equiv \Theta_K (x) \cup \Phi_K(x) \, . \label{familylimit1v}
\end{equation}
Of course, these conjunctions of kinks are congruent with the energy sum rules (\ref{eq:EnergySumRule}) as it is perceivable in Figure \ref{fig:1vDensidadesEnergia}. Note that also in this last case two different vrochosons arise in the same theory, which can be found at arbitrary separation distance as the kink family reveals.

\begin{figure}[ht]
\begin{center}
    \includegraphics[height=3.5cm]{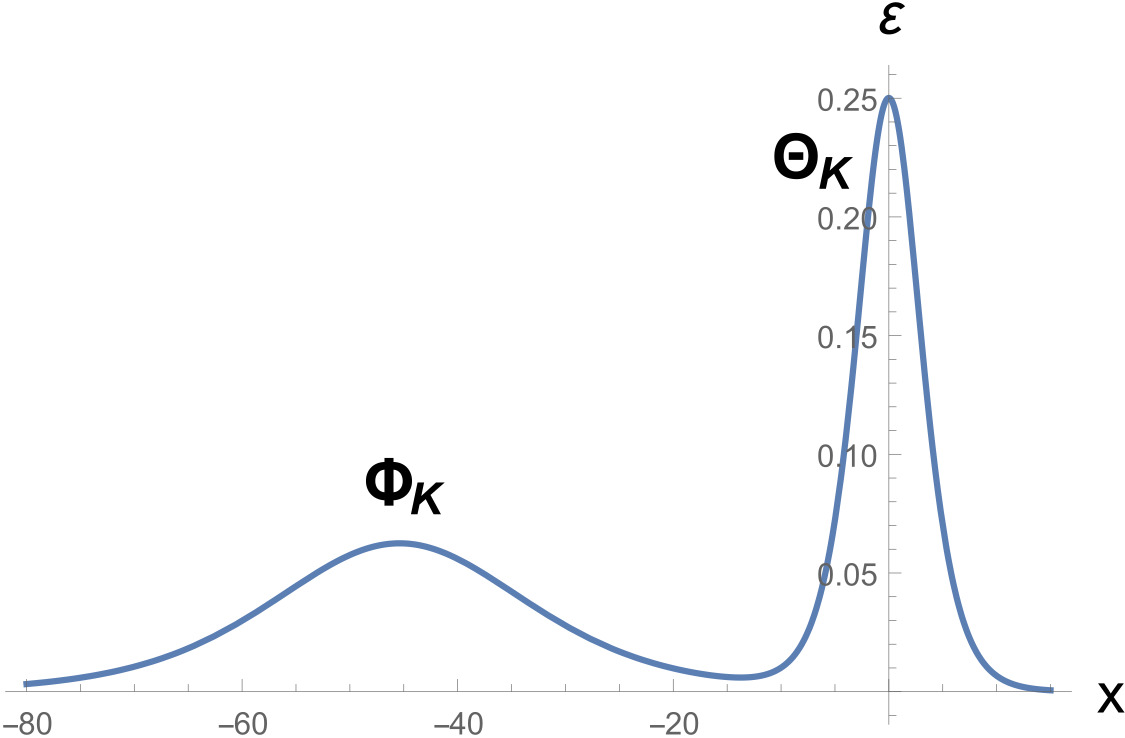} \hspace{0.3cm}
    \includegraphics[height=3.5cm]{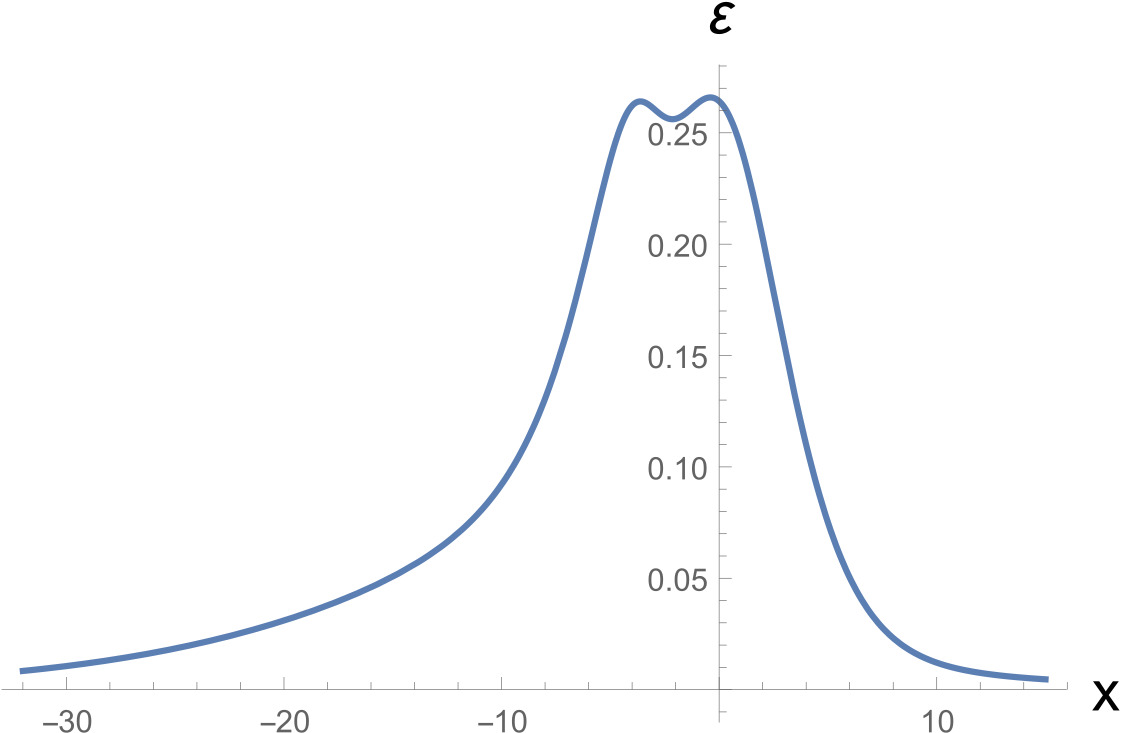} \hspace{0.3cm}
    \includegraphics[height=3.5cm]{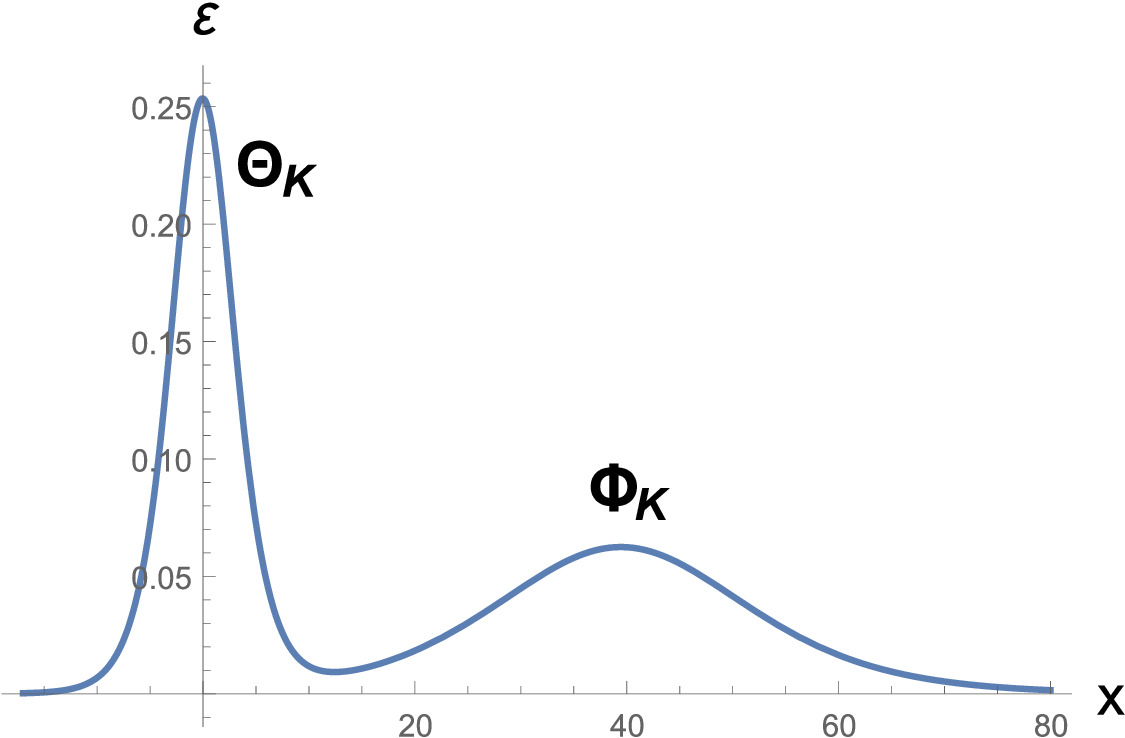}
    \caption{\small Energy densities for three members of the family $\Sigma_K^{(0,0,0)} (x;\gamma)$ given by $\gamma=-10$, $\gamma=0$ and $\gamma=10$ respectively. Values of parameters $(R,r,m_1,m_2)=(2,1,1,1)$ have been used. Notice how combinations of singular kinks are obtained as limits. }\label{fig:1vDensidadesEnergia}
    \end{center}
\end{figure}

These are non-contractible non-topological kinks and therefore these are also globally stable. In fact, this constitutes another example of globally stable families of non-topological kinks like \cite{A}. Furthermore, these solutions lack intermediate conjugate points, fact that makes them stable from the point of view of families of kinks. However, unlike the family of models in \cite{A}, this family of potential functions admits a model with only one vacuum point and therefore a scenario where all kinks are forced to be non-topological.
\end{itemize}

\section{Conclusions and further comments}

In this work a family of potentials for Sigma models on the torus $(\mathbb{S}^1\times \mathbb{S}^1)$ is constructed from a family of separable superpotentials such that a model with only one vacuum point can be found. The fact that the superpotential is separable has two direct consequences. The first one is there exist two non-trivial different superpotentials that minimize Bogomol'nyi energy. This may allow two different one-parameter families of kinks to emerge. The second one is that the superpotentials with which the Bogomol'nyi arrangement can be performed have the same minimum level of energy. This implies in turn that the stability of these kinks is ensured given the smoothness of the superpotential. Nevertheless, given the existence of families of kinks, two zero modes would be expected to appear in the spectral analysis of the Hessian operator for these solutions.

When the target manifold of a Sigma model is a simply connected manifold, homotopy classes of kinks are determined by the minima the kinks are connecting asymptotically. In that case a finite number of topological sectors arise depending on the number of minima of the model. However, when the target manifold is non-simply connected this may not be true anymore. In fact, for each two fixed minima linked by a kink on the torus, an infinite number of homotopy classes appear. In our case, that is the two-dimensional torus, winding numbers in both directions, toroidal and poloidal, give rise to infinite topological sectors. Nevertheless, the term topological cluster have been used to classify solutions according to the minima they are connecting.   

The kink variety for the particular models with four, two and only one minima is thoroughly studied. A different number of singular kinks, those with either $\theta$ or $\varphi$ constant, are obtained in each model. Their energies can be modulated by parameters $m_1$ and $m_2$. While the energy density profiles of all singular $\Theta_K-$kinks are indistinguishable in every model, $|n_1|$ and $|n_1|+1$ types of energy density profiles appear in each model for $\Phi_K-$kinks when $n_1$ is even and odd respectively. In particular, three types of singular kinks are obtained in the first two cases while only two in the last two. These extra profiles appear as a result of the asymmetry by construction between toroidal and poloidal coordinates. Indeed, exterior and interior $\Phi_K-$kinks in the first two cases, unlike $\Theta_K-$kinks, describe curves with different length. When only one toroidal singular kink can appear, as is the case with Cases $3$ and $4$,  these extra profiles disappear. 

Apart from these singular kinks, one-parameter families of kinks are found in each model, whose energies are simultaneously modulated by both $m_1$ and $m_2$. In fact, the energy of all members of these families comply with rules of sum with respect the energies of the singular kinks. Indeed, different limit members of these families coincide with combinations of singular kinks. 

On the other hand, given the symmetries that relate solutions of the Bogomol'nyi equations for the four different superpotentials for each potential, the torus is divided into different numbers of regions where the kink variety is replicated. In fact, the number of equivalent regions found in each case matches the number of minima $|4 n_1 n_2|$ of the potential function. This identification of solutions from different superpotentials affects both to singular and members of the families. However, only in the first case with four minima these transformations change the topological cluster of the solution. In particular, only topological kinks are found in the model of four minima (Case $1$), four singular kinks and two families of kinks replicated in the four regions of the torus. This was expected given that the superpotential is periodic on the torus in this case. However, models with two or one minima do admit non-topological kinks since the superpotential is not periodic on the torus. In Cases $2$ and $3$ with two minima topological and non-topological kinks coexist. Two topological and two non-topological kinks and one family of topological kinks are found in both regions of the torus. In Case $4$, with only one minimum and one region, all kinks are forced to be non-topological. Two non-topological singular kinks and one family of non-topological kinks are identified.

It is interesting to highlight that all these non-topological kinks are globally stable, since no kink is contractible to a point. Given this property, the term ``vrochoson'' has been coined to refer to these solutions. Lastly, it is of note that in all these models kinks revolve at most once around each poloidal and toroidal directions. It might be interesting to explore in the future different models where this scarcity of topological sectors is averted.

\end{document}